\documentclass[pra, twocolumn,showpacs,nofootinbibfloatfix,amsmath,amsfonts,amssymb]{revtex4-1}%
\usepackage{amsmath,amsfonts,amssymb,color}
\usepackage{amsthm}
\usepackage{leftidx}
\usepackage{graphicx}
\usepackage{xcolor}
\usepackage{cleveref}
\usepackage{colortbl}
\usepackage{tikz}
\usepackage{multirow}

\usepackage{subcaption} % Required for subfigures

\usepackage{bm}
\usepackage{epstopdf}
\usepackage{epsfig}
\usepackage{mathdots} 
\usepackage{caption}
\usepackage{graphicx}
\usepackage{subcaption}
\usepackage{braket}
\usepackage{amssymb}
\usepackage{mathrsfs}
\usepackage{cleveref}
\creflabelformat{equation}{(#2#1#3)}
\crefrangeformat{equation}{(#3#1#4--#5#2#6)}
\usepackage{amssymb}
\newcommand\Tau{\mathcal{T}}
\usepackage{chngcntr}
\usepackage{float}

\usepackage{lipsum} % for dummy text
\usepackage{subcaption}
\usepackage{placeins} % For \FloatBarrier
\usepackage{tikz}  % For drawing
\usepackage{array} % For advanced table features

\usepackage{multirow}  % For multirow functionality
\usepackage{array}     % For advanced table features
\usepackage{colortbl}  % For coloring cells
\usepackage[utf8]{inputenc}

\usepackage{titlesec}
\usepackage{lipsum} % for dummy text

\newcommand{\cmark}{\ding{51}} % Checkmark
\newcommand{\xmark}{\ding{55}} % Cross mark
\usepackage{colortbl}
\usepackage{pifont} % For additional symbols
\usepackage{mathtools}
\newenvironment{DATA AVAILABILITY}{%
  \par\noindent\textbf{Data Availability}\quad
}{\par}
\begin{document}
	\title{Anomalous topological edge modes in a periodically-driven trimer lattice} 
    
        \author{Mohammad Ghuneim}
        \email{mohammad.ghuneim@studio.unibo.it}
        \affiliation{%
		Department of Physics, King Fahd University of Petroleum and Minerals, 31261 Dhahran, Saudi Arabia
	}
	\author{Raditya Weda Bomantara}
	\email{Raditya.Bomantara@kfupm.edu.sa}
	\affiliation{%
		Department of Physics, Interdisciplinary Research Center for Intelligent Secure Systems, King Fahd University of Petroleum and Minerals, 31261 Dhahran, Saudi Arabia
	}
	\date{\today}
	
	%%%%%%%%%%%%%%%%%%%% ABSTRACT %%%%%%%%%%%%%%%%%%%%%%%%
	%\begin{linenumbers}
	
	\vspace{2cm}
	
\begin{abstract}
Periodically driven systems have a longstanding reputation for establishing rich topological phenomena beyond their static counterpart.  In this work, we propose and investigate a periodically driven extended Su-Schrieffer-Heeger model with three sites per unit cell, obtained by replacing the Pauli matrices with their $3\times 3$ counterparts. The system is found to support a number of edge modes over a range of parameter windows, some of which have no static counterparts. Among these edge modes, of particular interest are those which are pinned at a specific quasienergy value. Such quasienergy-fixed edge modes arise due to the interplay between topology and chiral symmetry, which are typically not expected in a three-band static model due to the presence of a bulk band at the only chiral-symmetric energy value, i.e., zero. In our time-periodic setting, another chiral-symmetric quasienergy value exists at half the driving frequency, which is not occupied by a bulk band and could then host chiral-symmetry-protected edge modes ($\pi$ modes). Finally, we verify the robustness of all edge modes against spatial disorder and briefly discuss the prospect of realizing our system in experiments.    
%Moreover, by applying periodic driving, we demonstrate the existence of robust $\pi$-edge modes as well as a rich edge state structure. Finally, photonic waveguide arrays and superconducting circuit platforms can be effectively used to implement our proposed scheme.

\end{abstract}

\maketitle

\section{Introduction}
 
Intriguing new phases of matter can be accomplished through the periodic driving of a quantum system. Primarily attributed to their knack for engendering exotic phases that often have no static counterparts, the fascination with driven systems has therefore sparked a frenzy of research efforts \cite{Slager2024, Chakrabarty2024, Sun2023, Zhu22, Weitenberg2021, R2021, Ikeda2020, Frederik2018}. Such periodically driven systems are often also referred to as Floquet systems. 
The dynamics generated by Floquet systems are determined by their quasienergies and the associated eigenstates, which are respectively analogous to energies and energy eigenstates in static systems. The quasienergy spectra of Floquet systems often exhibit nontrivial topological properties that have been studied in theoretical arenas \cite{Oka2009, Jiang2011, Rudner2013, R2016, Rahul2017, Zhou2018, Z2020, Harper2020, Huang24}, with recent experimental findings proving the existence of these features \cite{ Cheng2023, Wintersperger2020, Guglielmon2018, Mukherjee2018, Feis2025}. 

Floquet topological phases may arise in various physical platforms such as cold and ultracold atom systems \cite{Liu2019, Zhang2023}, photonic lattice \cite{Rechtsman2013, Afzal2020}, acoustic lattice \cite{Zhu22}, solid-state systems \cite{McIver2020}, and superconducting circuits \cite{Tan2020}. In particular, systems which are otherwise topologically trivial may become topologically nontrivial in the mere presence of appropriate periodic driving \cite{Lindner2011, Cayssol2013}. Moreover, in the presence of periodic driving, systems which are already nontrivial may become even more exotic and host a variety of new topological edge modes that have no static counterparts \cite{Jiang2011, Rudner2013, Quelle2017, B2021, Bomantara2022}. The unique topological features of Floquet topological phases have also been shown to find application in quantum computation \cite{R2020, Bomantara2020, Bomantara2018, R2018} and quantum simulation \cite{X2022}. 

%In systems devoid of edge states at first, introducing a time-periodic Hamiltonian can induce the formation of such states. However, the application of a time-dependent Hamiltonian can yield a deeper phase structure for systems that already include edge modes, allowing for the formation of new states or eliciting behaviors previously unavailable in systems with static counterparts \cite{Bomantara2019}. At its core, the periodic Hamiltonian causes a cyclic evolution that the system experiences, which can interpret the appearance of phases that are topologically nontrivial.

In the general field of topological phases, the Su-Schrieffer-Heeger (SSH) model \cite{Su1980} is one of the simplest toy models that supports nontrivial topological features. It describes a tight-binding chain with alternating (nearest-neighbor) hopping amplitudes, thereby forming two species of sublattice sites. In this case, depending on the ratio between the two hopping amplitudes, the system could either be in a topologically nontrivial regime, which supports a pair of topological edge modes or in a topologically trivial regime, which does not host such modes. Due to its simplicity, the topological characterization of the SSH model is analytically tractable. Moreover, while originally the SSH model describes a polyacetylene chain \cite{Su1980}, it has been experimentally realized on various platforms such as photonic/acoustic waveguides \cite{Klauck2021, LI2018}, mechanical systems \cite{Thatcher2022}, electrical circuits \cite{LIU2022, Splitthoff2024, Huang2024}, and quantum simulators \cite{Zhang2021}.

Despite its simplicity, the SSH model sufficiently demonstrates the distinction between characteristics of topologically trivial and non-trivial systems  \cite{Asboth2016}. As a result, the SSH model often forms the basis for more sophisticated topological systems involving additional bands \cite{Ghuneim2024, Anastasiadis2022, Alvarez2019, Verma2024, Du2024, Lee2022, Marques2020, M20218}, larger physical dimensions \cite{Benalcazar2017, Benalcazar17, Li2018, Bomantara2019}, and/or periodic driving \cite{J2013, J2014, Jangjan2022, Agrawal2022}. Such topological systems have also been subjects of studies in the areas of spintronics \cite{Keshtan2020, Wu2017}, quantum computing \cite{Koh2022, Petropoulos2022}, superconducting circuits \cite{Jin2023, Zheng2022, Deng2022}, topological photonics \cite{Tang2022, Chen2022}, spin-phonon systems \cite{Dong2022, Xiao2021}, quantum dots \cite{Maurer2022, Pham2022}, and machine learning \cite{Huembeli2018, Wong2023, Kairon2024}.  

With its simple but rich topology, a periodically driven SSH model is particularly attractive for investigating novel properties of Floquet topological phases in Hermitian \cite{Lago2015, Jangjan2020, Iadecola2024, Cheng2022} and non-Hermitian \cite{Shen2024, Vyas2021, Wu2020, Shi2024} systems. A key feature of a typical Floquet SSH model is the presence of the so-called $\pi$ edge modes, which are the generalizations of the zero edge modes in a static SSH model. In particular, while zero edge modes refer to edge localized eigenstates that are pinned at zero energy, $\pi$ edge modes are edge localized eigenstates that are pinned at half the driving frequency \cite{Tan2020, J2013, J2014}. That $\pi$ modes are unique to a Floquet SSH model and are attributed to the periodicity of quasienergy space by the driving frequency, which is in contrast to the unbounded nature of energy space in static systems. As $\pi$ and zero edge modes could coexist in a Floquet SSH model, the latter further serves as a promising platform for quantum technological applications \cite{Tan2020}.

Experimentally, $\pi$ edge modes have been demonstrated on numerous platforms. For instance, photonic lattice and waveguide arrays offer a fruitful ground for observing the $\pi$ edge modes due to their control over the driving parameters \cite{Cheng2019, Wu2021, Arkhipova2023, Wu2022}. These modes have also been unambiguously observed in solid-state systems where the quantum simulator has been used \cite{Chen2021}. Moreover, the Floquet $\pi$ modes are realized using plasmonic waveguide arrays, where evanescent coupling effectively simulates time-periodic hopping \cite{Sidorenko2022}.

In recent years, extensions to the SSH model with an enlarged unit cell, e.g., trimers \cite{Ghuneim2024, Du2024, Alvarez2019, Anastasiadis2022, Verma2024} and beyond \cite{Lee2022, Marques2020, M20218} have been considered. Such extended SSH models are generally found to yield significantly different characteristics from the regular SSH models. Indeed, some models break the conventional chiral symmetry \cite{Anastasiadis2022, Verma2024}, while retaining a modified form known as point chiral symmetry \cite{Anastasiadis2022}. Meanwhile, other extended SSH models, such as the one discussed in Ref.~\cite{Ghuneim2024}, preserve key symmetries including chiral, particle-hole, and time-reversal symmetry but nevertheless still exhibit distinct topological features from the regular SSH models. In particular, for the case of extended SSH models with three sites per unit cell, chiral symmetry (if it exists) forces a bulk band to have zero energy, thereby blocking the formation of topologically protected zero energy edge modes. In this case, the topology instead manifests itself as finite energy edge modes that are symmetrical about zero energy \cite{Ghuneim2024}. 

To the best of our knowledge, most existing studies on extended SSH models focus on the static setting. The interplay between periodic driving and the topology of extended SSH models is thus still largely exclusive. To fill in this knowledge gap, in this paper, we consider a periodically driven version of the extended SSH model with three sublattices per unit cell studied in Ref.~\cite{Ghuneim2024}. Our analysis below reveals the emergence of additional chiral-symmetric nonzero-energy edge states due to the presence of multiple topological phase transitions induced by the periodic driving. More remarkably, the sought-after $\pi$ edge modes are also found to exist in some cases. This finding is especially significant as it highlights another unique feature of Floquet systems that is otherwise unachievable in any static system. Indeed, constant energy edge states, such as the zero edge modes, are impossible to exist in chiral symmetric three-band systems due to the presence of a persisting zero energy bulk band that prohibits the formation of topological zero edge modes. While such a zero quasienergy bulk band also necessarily exists in our Floquet chiral-symmetric three-band system to similarly prevent the existence of topologically protected zero edge modes, the absence of a bulk band at half the driving frequency makes it possible to support $\pi$ edge modes. It is also worth noting that chiral symmetric nonzero energy edge states may coexist with the $\pi$ edge modes in some cases. This feature makes our proposed system fundamentally distinct from any static extended SSH model \cite{Ghuneim2024, Anastasiadis2022, Alvarez2019, Verma2024, Du2024, Lee2022, Marques2020, M20218} or Floquet regular (two-band) SSH model \cite{Tan2020, J2013, J2014, Jangjan2022, Agrawal2022}.

The rest of this paper is organized as follows: In Sec.~\ref{Floquet}, we present a concise overview of Floquet theory. In Sec.~\ref{model}, we provide the model description in real space. The momentum space analysis is presented in Sec.~\ref{mos}. In the same section, we also unveil the symmetries inherent to our model, identify the major band touching point locations, and define suitable topological invariants. In Sec.~\ref{res}, we display the quasienergy spectra in real space and demonstrate the formation of edge states. We briefly discuss potential experimental approaches for implementing our system in Sec.~\ref{Exp}. In Sec.~\ref{pert}, we consider four different types of perturbations and analyze their effects on the system. We further demonstrate the robustness of our system by investigating the effects of disorder in Sec.~\ref{dis}. Finally, we summarize our findings and present avenues for follow-up research in Sec.~\ref{conc}.

\section{Time-periodic extended SSH model}

\subsection{Floquet theory overview}
\label{Floquet}

Floquet theory is a very important tool for characterizing the physics of time-periodic systems. The theory amounts to solving the time-dependent Schr\"odinger equation, i.e., 
\[
i\frac{\partial}{\partial t} |\psi(t)\rangle = \mathcal{H}(t) |\psi(t)\rangle ,
\]
where $\hbar=1$ units are used throughout this paper and $\mathcal{H}(t)=\mathcal{H}(t+T)$ is the system's Hamiltonian that is periodic at period $T$. $|\psi(t)\rangle$ is a special class of states that satisfy the Floquet theorem
\[
|\psi(t)\rangle = e^{-{\mathrm{i}\varepsilon (t- t_0)}} |\psi(t_0)\rangle ,
\]
where $\varepsilon$ represents the quasienergy and $t_0$ is a reference initial time which will be set to $0$ in this paper. By denoting $U$ as the time-evolution operator over the course of one period, which will be referred to as \emph{the Floquet operator} onwards, an eigenvalue equation emerges as
\begin{equation}
    U |\psi(t_0)\rangle = e^{-\mathrm{i} \varepsilon T} |\psi(t_0)\rangle . \label{Floeig}
\end{equation}
Equation~\ref{Floeig} is analogous to the energy eigenvalue equation, the full solutions of which yield the system's quasienergy spectrum and its corresponding Floquet eigenstates. Despite its analogy with the concept of energy, quasienergy appears as a phase and is thus only defined up to a modulus of $2\pi/T$. In this paper, we restrict the quasienergy ``Brillouin Zone" to be within $[-\pi/T,\pi/T)$.

\subsection{Model description}
\label{model}

\begin{center}
\begin{figure}[htpb]
%\begin{figure}
  \includegraphics[width=0.45\textwidth]{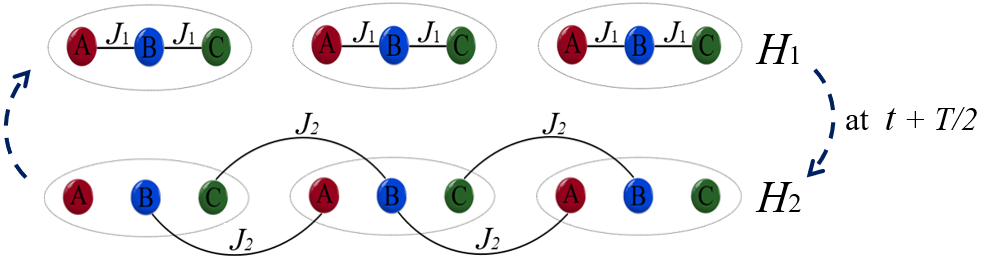}
  \captionof{figure}{The schematics of the system's Hamiltonian given in Eqs.~(\ref{eq:eq1})~and~(\ref{eq:eq2}) with $N=3$ unit cells. Each unit cell comprises three sublattices A, B, and C, represented by red, blue, and green, respectively.} 
  \label{fig:fig1}
\end{figure}    
\end{center}

We examine a periodically driven extended SSH model featuring three sites per unit cell, described by the two-time-step Hamiltonian ($n$ is the number of periods) 
\begin{flalign}
\mathcal{H}(t) &= \begin{cases}
    H_1 & \text{ for } n<t/T\leq (2n+1)/2 \\
    H_2 & \text{ for } (2n+1)/2<t/T \leq n+1 \\
\end{cases}
\label{eq:eq1}
\end{flalign}
where 
\begin{eqnarray}
    H_1 &=& \sum_{j=1}^{N}\, \left(J_{1}\,c_{A,j}^\dagger c_{B,j} + J_{1}\,c_{B,j}^\dagger c_{C,j} +h.c. \right) , \nonumber \\
    H_2 &=& \sum_{j=1}^{N-1}\, \left(J_{2}\,c_{A,j+1}^\dagger c_{B,j} + J_{2}\,c_{B,j+1}^\dagger c_{C,j} +h.c.\right),
    \label{eq:eq2}
\end{eqnarray}
$c_{\eta,j}^\dagger$  ($c_{\eta,j}$) is the creation (annihilation) operator at sublattice $\eta=A,B,C$ of the $j^\text{th}$ unit cell, $N$ is the number of unit cells, $T$ is the driving period, $J_{1}$ and $J_{2}$ are the intracell and  intercell  hopping parameters, respectively. Throughout this paper, we will consistently use  $N=25$ unit cells and a driving period of $T=2$ in our numerical calculations. The system's time-periodic Hamiltonian that comprises $H_1$ and $H_2$ is schematically depicted in Fig.~\ref{fig:fig1}. The intuition behind our model construction will be elaborated in Sec.~\ref{mos}.

The Floquet operator associated with Eq.~(\ref{eq:eq1}) reads
\begin{flalign}
    \hat{U} & = e^{(-i h_{2} T / 2)} e^{(-i h_{1} T / 2)},
    \label{eq:eq3}
\end{flalign}
where
\begin{eqnarray}
    h_{1} &=& \sum_{j=1}^N J_{1}  S_x \otimes |j\rangle \langle j|,  \nonumber \\ 
     h_{2} &=& \sum_{j=1}^{N-1} J_{2} S_+ \otimes |j\rangle \langle j+1| + h.c.,
\end{eqnarray}

$|j\rangle \equiv (0\cdots 1\cdots 0)^\dagger $, $S_+= \frac{1}{2} \left[S_x + i S_y\right]$, $S_x$ and $S_y$ are defined as follows:

\[
S_x = \left( \begin{smallmatrix} 
0 & 1 & 0 \\ 
1 & 0 & 1 \\ 
0 & 1 & 0 
\end{smallmatrix} \right), 
\quad
S_y = \left( \begin{smallmatrix} 
0 & -i & 0 \\ 
i & 0 & -i \\ 
0 & i & 0 
\end{smallmatrix} \right).
\]

\subsection{Momentum space analysis}
\label{mos}

The origin of our model construction could be better understood by writing Eq.~(\ref{eq:eq1}) in the momentum space, i.e.,
\begin{equation}
    \mathcal{H}(t)=H(k,t)\otimes |k\rangle \langle k | ,
\end{equation}
where periodic boundary conditions (PBC) are assumed, 
\begin{flalign}
H(k, t) &= \begin{cases}
    h_{1}(k) & \text{ for } n<t/T\leq (2n+1)/2 \\
    h_{2}(k) & \text{ for } (2n+1)/2<t/T \leq n+1 \\
\end{cases} ,
\end{flalign}
and
\begin{eqnarray}
h_{1}(k) &=& J_{1}\,S_{x},\nonumber \\
h_{2}(k) &=& J_{2} \cos(k)\,S_{x} + J_{2} \sin(k)\,S_{y}.
\label{eq:eq6}
\end{eqnarray}
The corresponding momentum space Floquet operator is then found as,
\begin{flalign}
    U(k) & = e^{(-i h_{2}(k) T / 2)}\,e^{(-i h_{1}(k) T / 2)} . 
    \label{eq:eq7}
\end{flalign}

The momentum space description is particularly useful for identifying the system's symmetries. For time-periodic systems, the chiral, time-reversal, and particle-hole symmetries are respectively identified as~\cite{Roy2017},
\begin{eqnarray}
    \Gamma h(\mathbf{k}, t) \Gamma^{-1} &=& -h(\mathbf{k}, T-t) , \nonumber \\
    \mathcal{T} h(\mathbf{k}, t) \mathcal{T}^{-1} &=& h(-\mathbf{k}, T-t) , \nonumber \\
    \mathcal{P} h(\mathbf{k}, t) \mathcal{P}^{-1} &=& -h(-\mathbf{k}, t) , \label{Flosym}
\end{eqnarray}
where $h(\mathbf{k}, t)$ is the momentum space Hamiltonian, $\Gamma$ is a unitary operator, whilst $\mathcal{T}$ and $\mathcal{P}$ are anti-unitary operators. The symmetries of our system become visible when its momentum space Hamiltonian is written at a shifted initial time such that  
%\begin{flalign}
%    \hat{U} & = e^{(-i h_{1}(k,t) T / 4)} e^{(-i h_{2}(k,t) T / 2)} e^{(-i h_{1}(k,t) T / 4)},
%    \label{eq:eq}
%\end{flalign}
\begin{flalign}
\tilde{H}_1(k, t) &= \begin{cases}
    h_{1}(k) &  0<t\leq \frac{1}{4}T \\
    h_{2}(k) &  \frac{1}{4}T<t \leq \frac{3}{4}T \\
    h_{1}(k) &  \frac{3}{4}T<t\leq T \\
\end{cases}.
\label{eq:eq}
\end{flalign}
We find that $\tilde{H}_1(k,t)$ satisfies all the symmetries described by Eq.~(\ref{Flosym}) with respect to $\Gamma=\text{diag}(-1, 1, -1)$, $\Tau=\mathcal{K}$ ($\mathcal{K}$ being the complex conjugation operator), and $\mathcal{P}=\Gamma \mathcal{K}$. Note that the same symmetry operators also apply when the momentum space Hamiltonian is written at another shifted initial time
\begin{flalign}
\tilde{H}_2(k, t) &= \begin{cases}
    h_{2}(k) &  0<t\leq \frac{1}{4}T \\
    h_{1}(k) &  \frac{1}{4}T<t \leq \frac{3}{4}T \\
    h_{2}(k) &  \frac{3}{4}T<t\leq T \\
\end{cases}.
\label{eq:eqb}
\end{flalign}

\begin{figure}[htpb]
    \begin{subfigure}[l]{0.14\textwidth}  \includegraphics[width=\linewidth]{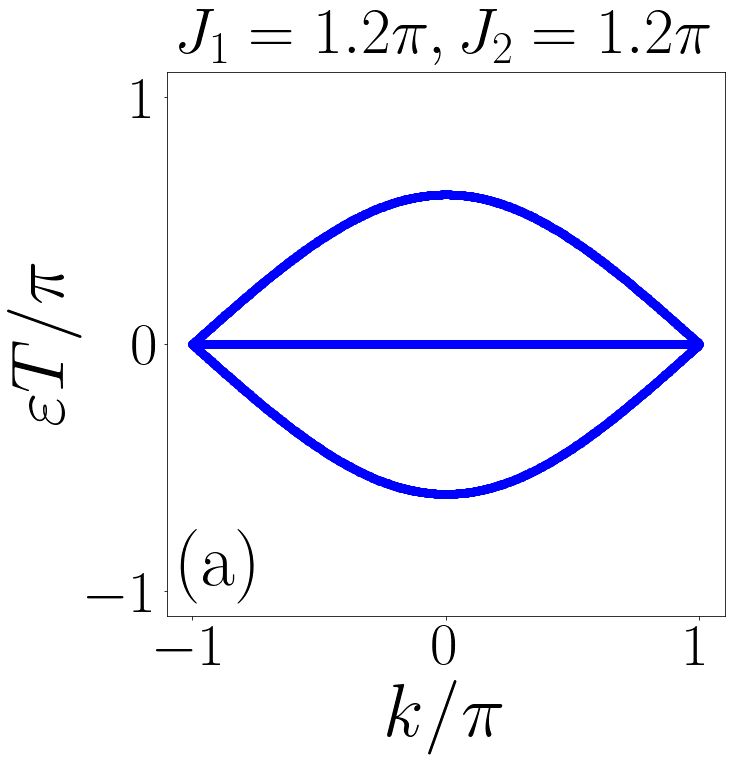}     
    \end{subfigure}%
    \hspace{-0.1cm} % Reduces space between subfigures
    \begin{subfigure}[c]{0.125\textwidth}
    \includegraphics[width=\linewidth]{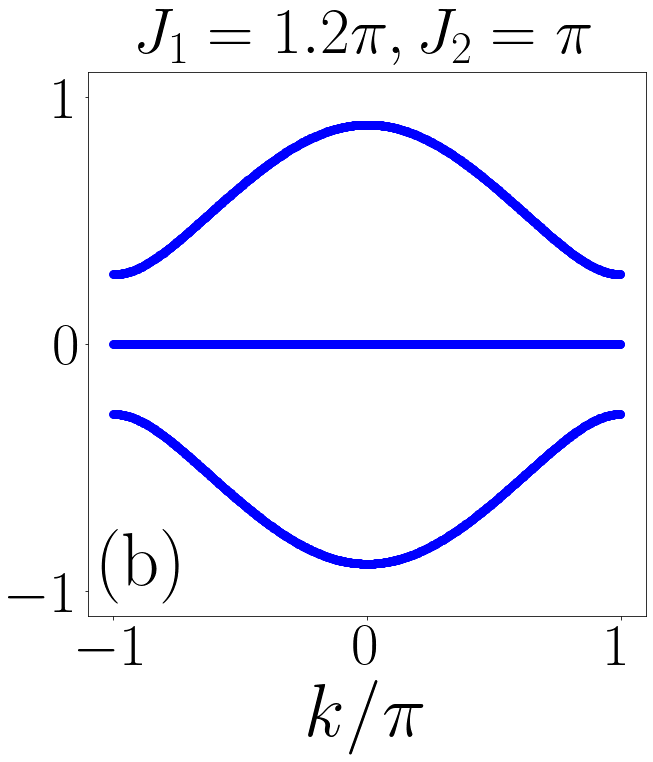}
    \end{subfigure}%
    \hspace{-0.1cm} 
    \begin{subfigure}[r]{0.125\textwidth}
      \includegraphics[width=\linewidth]{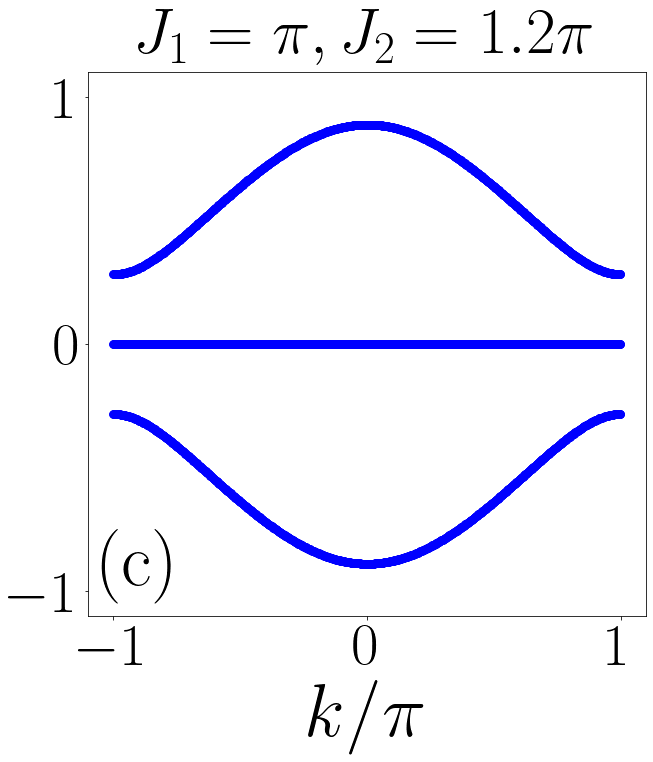}        
    \end{subfigure}
    \caption{Some representative quasienergy spectra corresponding to Eq.~(\ref{eq:eq12}) and Eq.~(\ref{eq:eq13}).}
    \label{fig:fig2a}
\end{figure}

The Floquet operators associated with the two momentum space Hamiltonians above can be explicitly written as
\begin{align}
U_1 &= e^{-\frac{i h_1(k)}{2} T/2} e^{-i h_2(k) T/2} e^{-\frac{i h_1(k)}{2} T/2} \label{eq:eq12}, \\
U_2 &= e^{-\frac{i h_2(k)}{2} T/2} e^{-i h_1(k) T/2} e^{-\frac{i h_2(k)}{2} T/2} \label{eq:eq13}.
\end{align}
As both operators are related by a unitary transformation, they share the same quasienergy spectrum, as shown in Fig.~\ref{fig:fig2a} for some representative parameter values. Note that while Fig.~\ref{fig:fig2a}(b,c) look qualitatively the same, they are topologically distinct as a topological phase transition occurs at $J_1=J_2$ (Fig.~\ref{fig:fig2a}(a)). Unlike in the static case where $J_1=J_2$ is the only topological phase transition point, we find that additional topological phase transitions occur as the parameters $J_1$ and $J_2$ are varied. Intuitively, such topological phase transitions are made possible by the bounded nature of the quasienergy Brillouin Zone, thereby allowing the two outer bands to meet the zero quasienergy line from two different directions. To explicitly identify these additional band-touching points, we first define the effective Floquet Hamiltonians
\begin{align}
\mathcal{H}_{\text{eff},1} 
 &\equiv \frac{U_1^\dagger - U_1}{2i} \label{eq:eq14}, \\
\mathcal{H}_{\text{eff},2}
 &\equiv \frac{U_2^\dagger - U_2}{2i} \label{eq:eq15}.
\end{align}
It follows that the eigenvalues $E$ of $\mathcal{H}_{\text{eff},1}$ and $\mathcal{H}_{\text{eff},2}$ are related to the quasienergies $\varepsilon$ by $E=\sin(\varepsilon T)$. As band touching points occur at quasienergy of either $0$ or $\pi/T$ in the presence of chiral symmetry, these correspond to all three eigenvalues of $\mathcal{H}_{\text{eff},1}$ and $\mathcal{H}_{\text{eff},2}$ being $0$. We further find that both $\mathcal{H}_{\text{eff},1}$ and $\mathcal{H}_{\text{eff},2}$ have the same structure of 
\begin{equation}
    \left(\begin{array}{ccc}
        0 & A(k) & 0  \\
        A^*(k) & 0 & A(k) \\
        0 & A^*(k) & 0 \\
    \end{array}\right) ,
\end{equation}
the three eigenvalues of which are identified as $E_0=0$ and $E_\pm=\pm \sqrt{2} |A|$. While the exact expression for $A(k)$ is generally quite complicated, upon fixing $k$ to either $0$ or $\pi$, the equation $E_\pm=0$ yields 
\begin{eqnarray}
    J_1 &=& \frac{n \pi}{\sqrt{2}} - J_2, \quad \text{at } k = 0 \nonumber , \\ 
J_1 &=& J_2 + \frac{m \pi}{\sqrt{2}}, \quad \text{at } k = \pi
\label{0pi} ,
\end{eqnarray}
where $m$ and $n$ are integers. For $m=n=0$, the above conditions reduce to the expected $J_1=\pm J_2$. It should also be emphasized that there may be additional band touching points not captured by the above relation, which correspond to $k\neq 0,\pi$. 

Finally, we will now define a set of potential topological invariants for our system. To this end, we adapt the winding numbers defined in Refs.~\cite{J2013, J2014}. In particular, Refs.~\cite{J2013, J2014} consider a two-band periodically driven system protected by a chiral symmetry, the topology of which manifests itself as zero and/or $\pi$ edge modes. A pair of integer-valued winding numbers, $\nu_0$ and $\nu_\pi$, could then be introduced to count the number of zero and $\pi$ edge modes, respectively. These invariants are defined, respectively, from the sum and difference between the winding numbers associated with some subblock of two effective Hamiltonians. Specifically, given a pair of two-band effective Floquet Hamiltonians corresponding to two different time-shifts, i.e., 
\begin{equation}
    h_{{\rm eff},j} = \left(\begin{array}{cc}
        0 & a_j \\
        a_j^* & 0 \\
    \end{array} \right)
\end{equation}
for $j=1,2$, the number of pairs of zero and $\pi$ edge modes are dictated by the two winding numbers
\begin{eqnarray}
    \nu_0 &=& \frac{\nu_1+\nu_2}{2} , \nonumber \\
    \nu_\pi &=& \frac{\nu_1-\nu_2}{2} , \nonumber \\
    \nu_j &=& \frac{1}{2\pi i} \int a_j^{-1} da_j . \label{w2b}
\end{eqnarray}
The above procedure can be easily generalized to a family of chiral symmetric periodically driven systems with a larger but even number of bands. In this case, $a_j$ becomes a square $M/2\times M/2$ matrix ($M$ being the number of bands), and the trace is taken in Eq.~(\ref{w2b}) to ensure that the resulting quantity is a number. This generalized set of winding numbers has indeed accurately described the number of topological edge modes in various systems \cite{Bomantara2022, R2018, R2020, Koor2022}.

As our system has three bands, it is impossible to find a suitable subblock that is a square matrix, thereby preventing the direct construction of the winding numbers above, i.e., Eq.~(\ref{w2b}) and their larger-band generalizations. In this case, our strategy is to modify the above procedure by replacing the problematic calculation of winding numbers with the calculation of Zak phases instead. Specifically, we define our two topological invariants as
\begin{eqnarray}
    \Phi_0^{(s)} &=& \frac{\chi_1^{(s)}+\chi_2^{(s)}}{2\pi} , \label{eq:eq16} \\ 
    \Phi_\pi^{(s)} &=& \frac{\chi_1^{(s)}-\chi_2^{(s)}}{2\pi} \label{eq:eq17} .
\end{eqnarray}
where $\chi_1^{(s)}$ and $\chi_2^{(s)}$ are the Zak phases 
\begin{equation}
    \chi_j^{(s)} = i \int_{-\pi}^{\pi} \langle E_{s,j}(k) | \partial_k | E_{s,j}(k) \rangle dk , \label{Zaks}
\end{equation}
$|E_{s,j}(k)\rangle$ is the energy eigenstate associated with the $s$th band of $\mathcal{H}_{\text{eff},j}$ ($j=1,2$) defined in Eqs.~(\ref{eq:eq14}) and (\ref{eq:eq15}).

Our construction above is motivated by the fact that, for a two-band chiral symmetric system, the Zak phase is related to the winding number by $\nu_j = \frac{\chi_j}{\pi}$. Consequently, Eqs.~(\ref{eq:eq16}) and (\ref{eq:eq17}) are equivalent to those defined in Refs.~\cite{J2013, J2014} in a two-band chiral symmetric periodically driven system. While this equivalence may no longer hold for systems with a larger number of bands, it is expected that such Zak phases may still carry some topological meaning. Therefore, it should be noted that the subscripts $0$ and $\pi$ in Eqs.~(\ref{eq:eq16}) and (\ref{eq:eq17}) are made merely for consistency with the notations of Refs.~\cite{J2013, J2014}, but $\Phi_0^{(s)}$ and $\Phi_\pi^{(s)}$ may not strictly correspond to the presence of zero and $\pi$ edge modes. Indeed, despite the absence of zero edge modes in our three-band system, we find that $\Phi_0^{(s)}$ may still be non-zero if other types of edge states are present. We will present the full numerical calculation and analysis of $\Phi_0^{(s)}$ and $\Phi_\pi^{(s)}$ in the next section so that a direct comparison with the presence of system edge states can be made.

\subsection{Real space analysis}
\label{res}
\begin{center}
\begin{figure}[htpb]
    \centering

    \begin{subfigure}{0.45\textwidth}
       \includegraphics[width=\linewidth]{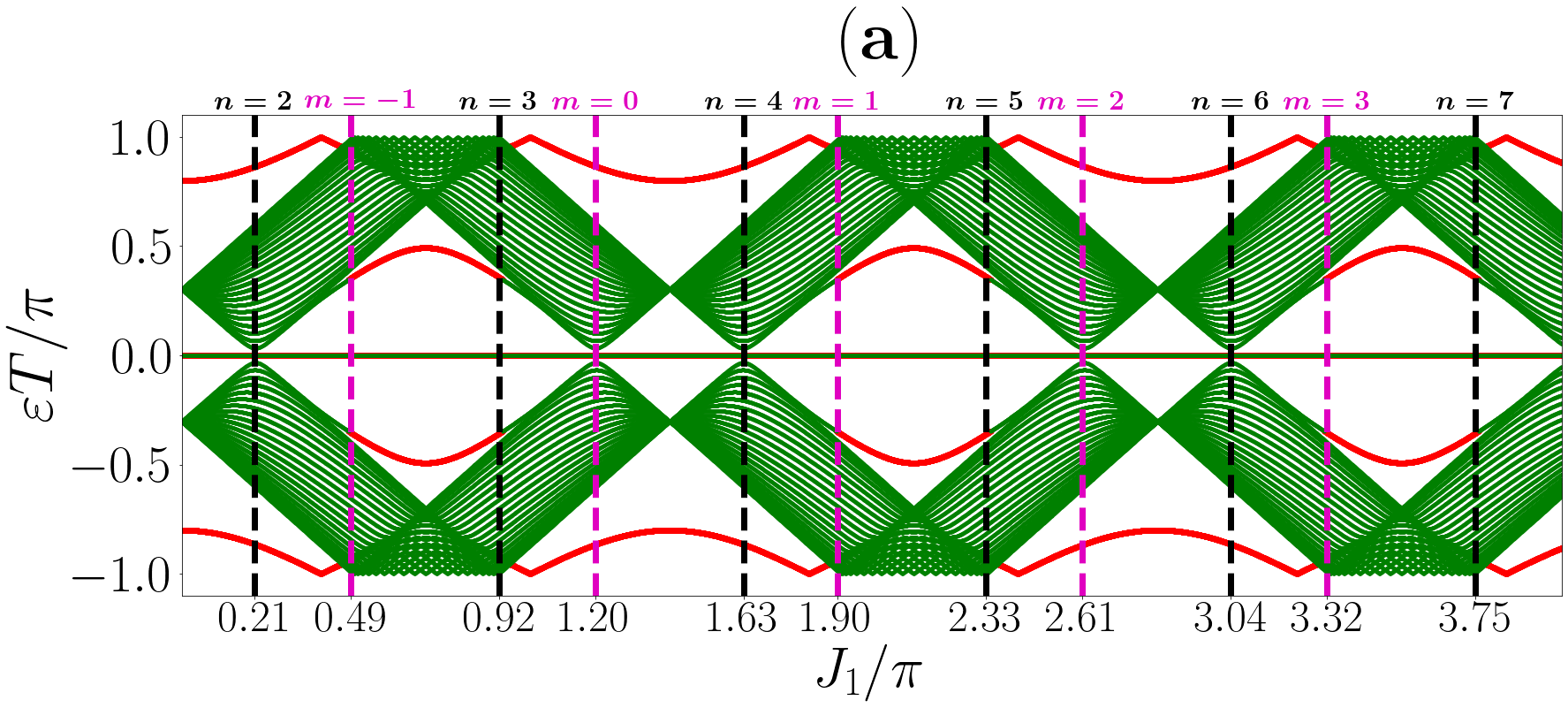}
    \end{subfigure}
    
    \vspace{0.1cm} 
    \begin{subfigure}{0.45\textwidth}
        \includegraphics[width=\linewidth]{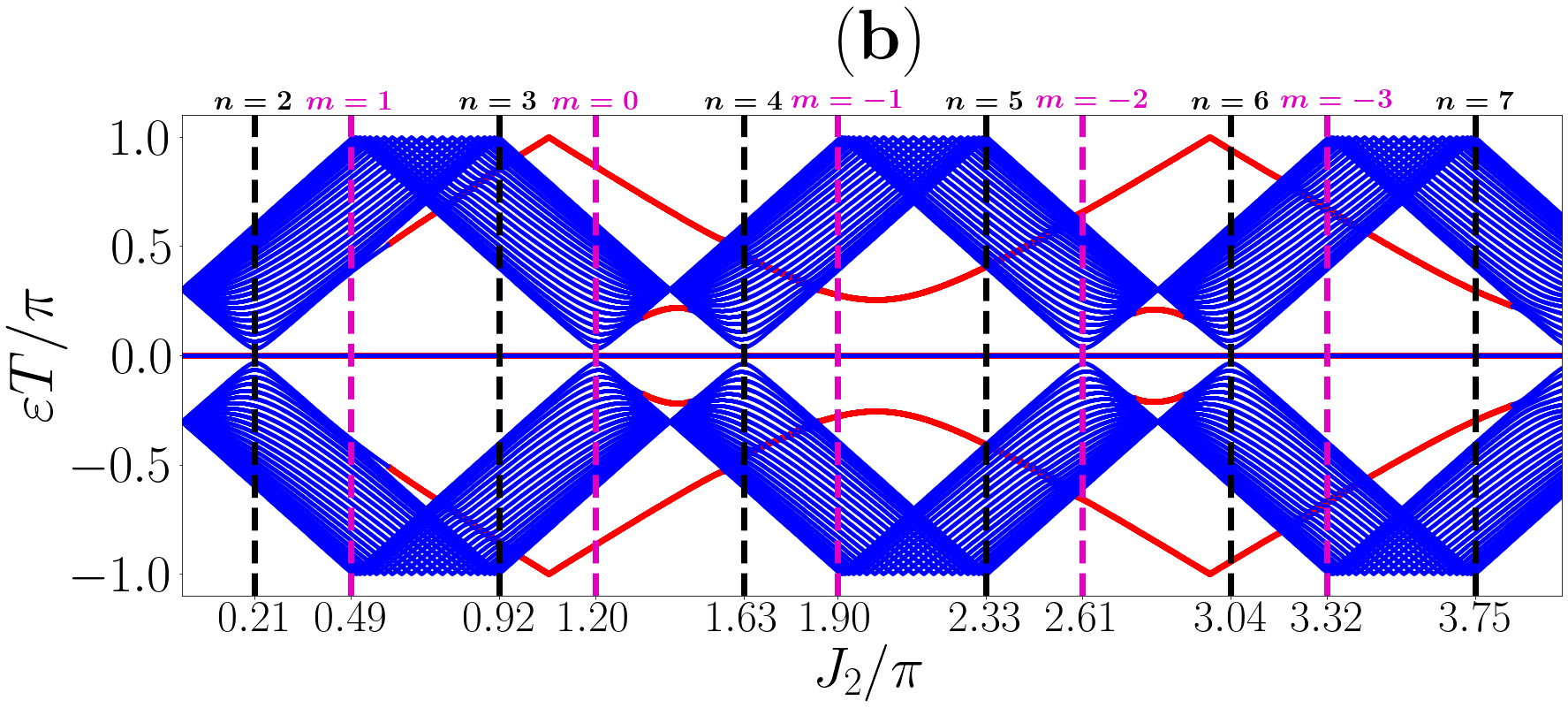}
    \end{subfigure} 
    \caption{The real space quasienergy spectrum associated with Eq.~(\ref{eq:eq3}) for $N=25$ unit cells and $T=2$. (a) is fixed at $J_{2}=1.2\pi$ and (b) is fixed at $J_{1}=1.2\pi$. The system's edge modes are represented in red. The black and pink vertical dashed lines mark the locations of the band touching points predicted by Eq.~(\ref{0pi}).}
    \label{fig:fignm}
\end{figure}
\end{center}

\begin{figure}[htpb]
    \centering
    \begin{subfigure}[b]{0.45\textwidth}
        \includegraphics[width=\textwidth]{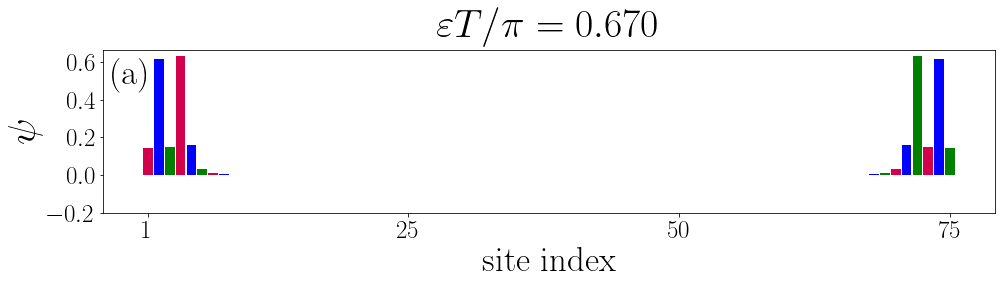} 
        \label{fig:fig1a}
    \end{subfigure}
    \begin{subfigure}[b]{0.45\textwidth}
        \includegraphics[width=\textwidth]{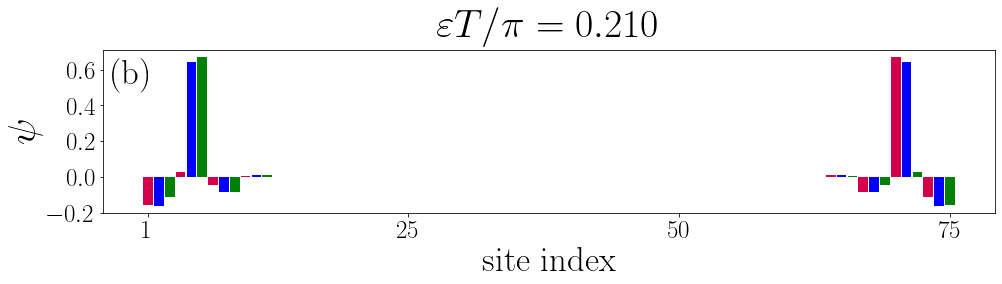}
        \label{fig:fig1b}
    \end{subfigure}
    \begin{subfigure}[b]{0.45\textwidth}
        \includegraphics[width=\textwidth]{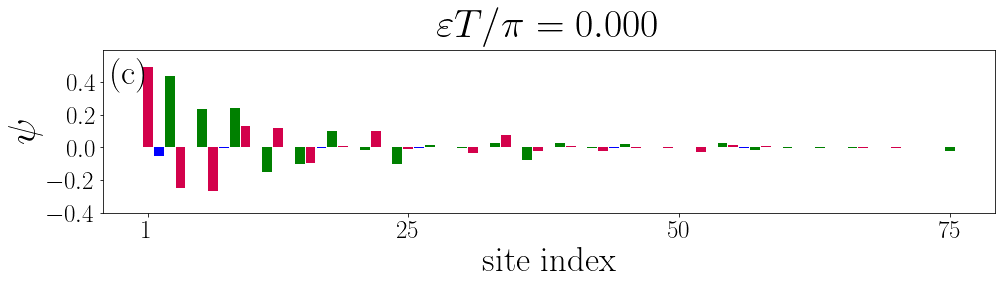} 
        \label{fig:fig1c}
    \end{subfigure}
    \begin{subfigure}[b]{0.45\textwidth}
        \includegraphics[width=\textwidth]{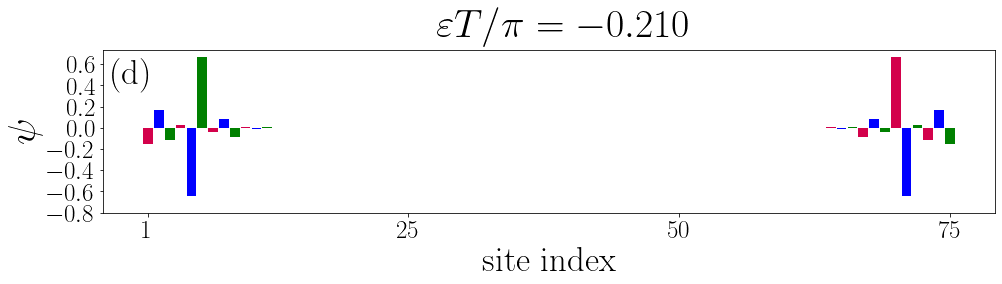}
        \label{fig:fig1d}
    \end{subfigure}
    \begin{subfigure}[b]{0.45\textwidth}
        \includegraphics[width=\textwidth]{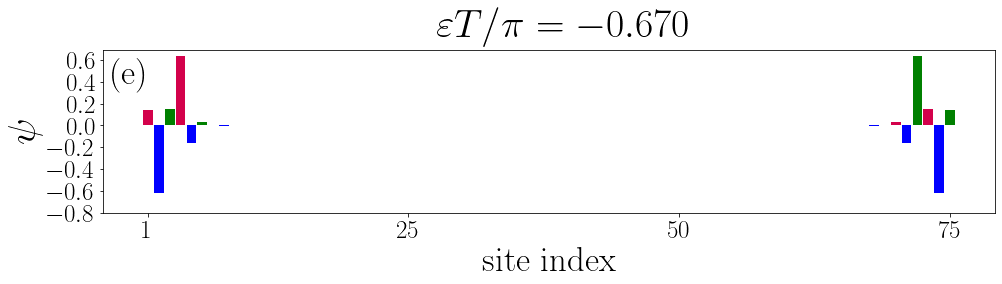}
        
        \label{fig:fig1e}
    \end{subfigure}
    \caption{Wave function profiles associated with the various edge states of the system with $J_{1}= 1.2\pi$, $J_{2}=1.4\pi$, $N=25$ unit cells, and $T=2$ (along the orange solid line in Fig.~\ref{fig:fig2}(b)). Notice that a pair of edge state profiles (one left-localized and one right-localized) are plotted together in all panels except for the case of $\varepsilon T/\pi = 0$, where only one left-localized edge state exists.}
    \label{fig:fig3}
\end{figure}

To reveal the formation of edge states in our system, we now turn our attention back to the system's real-space description under open boundary conditions (OBC), i.e., Eqs.~(\ref{eq:eq1}) and (\ref{eq:eq2}). In Fig.~\ref{fig:fignm}, we show the corresponding quasienergy spectrum of the system as a function of one hopping parameter while the other hopping parameter is fixed. There, we observe the multiple occurrences of band touching points, which exactly match the predictions of Eq.~(\ref{0pi}). Moreover, such band touching points are accompanied by the emergence and disappearance of nonzero quasienergy edge states, further suggesting that they correspond to topological phase transitions. In Fig.~\ref{fig:fig3}, the wave function profiles associated with the system's edge states at a fixed set of parameter values are computed. It follows that all edge states at non-zero quasienergy values come in pairs, one left-localized and the other right-localized. Interestingly, we also observe a single left-localized edge state at zero quasienergy. However, it should be noted that such an edge state is not topological, as it coexists with the system's zero quasienergy bulk band. As compared with the other edge states, this zero quasienergy edge state also appears to have a much larger localization length.

Among the various edge states in the system, it is observed that those that exist within the two bulk bands (e.g., between $J_1=0.2\pi$ and $J_1=1.2\pi$ in Fig.~\ref{fig:fignm}(a)) are analogous to the edge states observed in the static counterpart of the system as previously studied in Ref.~\cite{Ghuneim2024}. It is worth noting that in the static extended SSH model of Ref.~\cite{Ghuneim2024}, such edge states are observed for $J_1<J_2$. On the other hand, in the present time-periodic setting, such edge states appear and disappear periodically (following the gap closing between the two bulk bands as predicted in Eq.~\ref{0pi}) as one hopping parameter is varied while the other is fixed. This significant distinction between the edge states' behavior in the static and the time-periodic systems is attributed to the fact that $J_1$ and $J_2$ appear as arguments of a complex exponential function in the Floquet operator of Eq.~(\ref{eq:eq3}), thus explaining their periodic behavior in our time-periodic system. 

In addition to the above type of edge states that exist within the two bulk bands, we also observe a different type of edge states that exist beyond the two bulk bands (e.g., between $J_1=0$ and $J_1=0.49\pi$ in Fig.~\ref{fig:fignm}(a)). In Fig.~\ref{fig:fignm}(a), such edge states persist throughout the parameter values under consideration and do not appear to be affected by band touching points. However, Fig.~\ref{fig:fignm}(b) reveals that such edge states originate from a band touching point that arises at a smaller value of $J_2$, i.e., at around $J_2\approx 0.21\pi$ when $J_1=1.2\pi$. Interestingly, once such edge states emerge, it persists for any value of $J_1$ despite the periodic presence of band touching points as $J_1$ is varied, which could be attributed to the fact that, unlike the edge states that exist within the bulk bands, they do not merge with the bulk bands as a band touching point occurs, i.e., these edge states have negligible overlap with the bulk eigenstates. Moreover, due to the periodicity of the quasienergy Brillouin zone, the positive and negative pairs of these edge states periodically meet at $\pm \pi/T$ quasienergy. These two features deem this second type of edge states unique to our periodically driven system and have no static counterpart. 

It is worth noting that the two types of non-zero quasienergy edge states identified above, i.e. those that, respectively, develop within and beyond the two bulk bands, have different structures on the lattice sites. For example, as demonstrated in Fig.~\ref{fig:fig3}, while the left-localized edge states that exist within the two bulk bands (at quasienergy around $\pm 0.210\pi$) have the largest support on sublattice B and C of the second unit cell, the left-localized edge states that exist beyond the two bulk bands (at quasienergy around $\pm 0.670\pi$) have the largest support on sublattice B of the first unit cell and sublattice A of the second unit cell. A similar distinction is also observed between the two types of edge states that are localized to the right edge.
\begin{center}
    \begin{figure*}[htpb]
    \begin{subfigure}[l]{1\textwidth}
        \includegraphics[width=\linewidth]{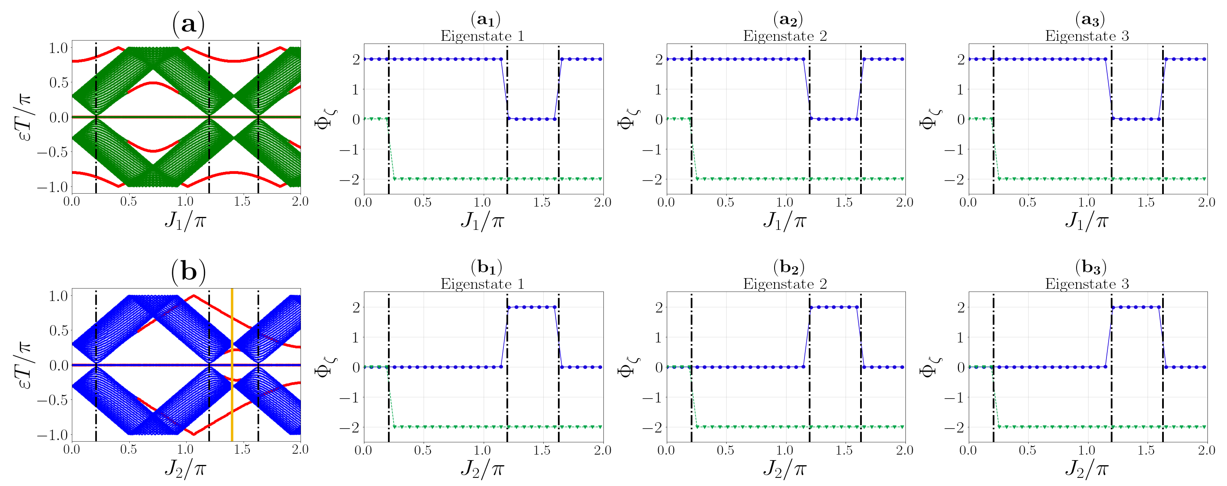}
    \end{subfigure}%  
    \caption{Subplots (a) and (b) represent the quasienergy spectrum associated with Eq.~(\ref{eq:eq3}) at $N=25$ unit cells and $T=2$. (a) is fixed at $J_{2}=1.2\pi$ and (b) is fixed at $J_{1}=1.2\pi$. The edge modes are represented in red. The orange solid vertical
    line on subplot (b) marks the parameter values used in Fig.~\ref{fig:fig3}.  Subplots $(a_{1})\text{-}(a_{3})$ and $(b_{1})\text{-}(b_{3})$, respectively, describe the corresponding topological invariants \raisebox{-0.1cm}{\includegraphics[width=1.5cm]{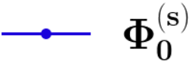}} and \raisebox{-0.1cm}{\includegraphics[width=1.5cm]{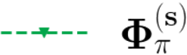}} for each eigenstate $s=1,2,3$. The black dashed-dotted vertical lines in all subplots are positioned at identical values, indicating the points where phase transitions occur.}
    \label{fig:fig2}
\end{figure*}
\end{center}

We close this section by presenting the numerically calculated topological invariants $\Phi_0^{(s)}$ and $\Phi_\pi^{(s)}$ (see Eqs.~(\ref{eq:eq16}) and (\ref{eq:eq17}), respectively) and associating them with the formation of edge states. Our results are summarized in Fig.~\ref{fig:fig2}. There, we indeed confirm that the band touching points that lead to the change in the number of edge states are accompanied by a jump in either $\Phi_0^{(s)}$ or $\Phi_\pi^{(s)}$ regardless of the eigenstate index $s$. 

One may wonder how $\Phi_0^{(s)}$ and $\Phi_\pi^{(s)}$, which traditionally quantify the number of zero edge modes and $\pi$ edge modes respectively \cite{J2013, J2014}, could be non-zero even though there is no zero edge mode or $\pi$ edge mode in the system. The answer lies in the fact that, strictly speaking, $\Phi_0^{(s)}$ and $\Phi_\pi^{(s)}$ should only be defined up to a modulus of $2$ as they are proportional to Zak phases. In this case, a value of $1$ ($0$) modulo 2 corresponds to the presence (absence) of edge modes. Since the calculated $\Phi_0^{(s)}$ and $\Phi_\pi^{(s)}$ are even, i.e., $0$ modulo 2, everywhere in Fig.~\ref{fig:fig2}, they are consistent with the absence of zero edge mode and $\pi$ edge mode. However, the fact that the non-modulus values of $\Phi_0^{(s)}$ and $\Phi_\pi^{(s)}$ exhibit jumps in the presence of band touching points suggests that they could also be utilized to capture edge modes beyond zero edge modes and $\pi$ edge modes, as clearly seen in Fig.~\ref{fig:fig2}. 

Upon comparing the leftmost panel (e.g., Fig.~\ref{fig:fig2}(a)) with any panel to its right (e.g., Fig.~\ref{fig:fig2}($\text{a}_1$)), we observe that the whole values of $\Phi_0^{(s)}$ and $\Phi_\pi^{(s)}$ collectively determine all types of edge modes in the system. In particular, notice that both $\Phi_0^{(s)}$ and $\Phi_\pi^{(s)}$ are nonzero when both types of edge modes that exist inside and outside bulk bands exist (e.g., between $J_1=0.21\pi$ and $J_1=1.2\pi$). Meanwhile, only one of $\Phi_0^{(s)}$ and $\Phi_\pi^{(s)}$ is nonzero when only one type of edge modes is present (e.g., between $J_1=1.2\pi$ and $J_1=1.63\pi$). Finally, both $\Phi_0^{(s)}$ and $\Phi_\pi^{(s)}$ are zero when there is no edge mode (e.g., between $J_2=0$ and $J_2=0.21\pi$ in Figs.~5(b) and 5($\text{b}_1$)). While an alternative set of topological invariants that characterize each type of edge modes individually may exist, their construction unfortunately poses a nontrivial task and deserves a separate study. Nevertheless, given that all types of edge modes are in fact already accounted for by $\Phi_0^{(s)}$ and $\Phi_\pi^{(s)}$, our proposed invariants are sufficient to confirm their topological nature.

\section{Discussion}
\subsection{Experimental realization}
\label{Exp}
The one-dimensional SSH model and its variations are among the simplest topological models to implement in the laboratory. One promising approach is through the use of photonic waveguides, where each waveguide represents a lattice site and the coupling between them simulates hopping parameters in the Hamiltonian \cite{Yang2024, Aravena2022}. In this case, Floquet driving can be achieved by applying periodic modulations to the waveguide structure, allowing the system to evolve under time-dependent conditions \cite{Cheng2019, Wu2021, Arkhipova2023, Wu2022}.

Additionally, superconducting circuits provide a platform where the extended SSH model can be realized \cite{Youssefi2022, Cai2019, Splitthoff2024}. Qubits can represent lattice sites, and the tunable couplings between them can simulate the hopping terms \cite{Mei2018}. Floquet driving is performed by applying external time-periodic fields to modulate the system’s parameters. Recent advancements in quantum simulators based on superconducting qubits provide high precision in controlling the system dynamics, making them excellent candidates for realizing such one-dimensional topological models \cite{Chen2021}.

\begin{center}
\begin{figure}[htpb]
%\begin{figure}
  \includegraphics[width=0.4\textwidth]{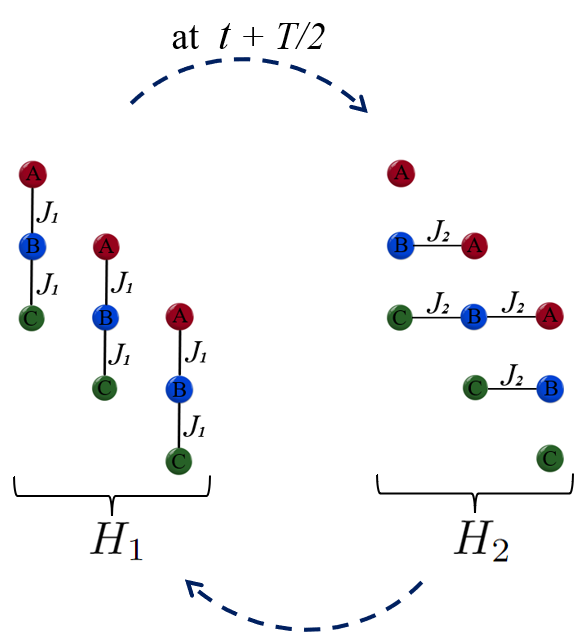}
  \captionof{figure}{The schematic representation of our model in a rearranged lattice that turns all hopping parameters nearest-neighbor.} 
  \label{fig:figH12}
\end{figure}    
\end{center}

To circumvent the difficulty of realizing the long-range hopping that is present in $H_2$, we may rearrange the lattice into a ladder-like structure as shown in Fig.~\ref{fig:figH12}. This way, all hoppings in our time-periodic Hamiltonian become strictly nearest-neighbors in nature, which is easy to achieve. Note that the lattice arrangement of Fig.~\ref{fig:figH12} could either consist of waveguides or superconducting circuits, i.e., the design is compatible with both experimental platforms.

\subsection{Effects of perturbation}
\label{pert}

To investigate the fate of the observed edge states, as well as to further uncover the interplay between symmetries and topology in our system, we will now consider the presence of some representative perturbations. Specifically, we consider the following perturbations that in the momentum space read,

\begin{eqnarray}
    \alpha(k) \equiv \alpha_{\text{o}} \scalebox{0.6}{$\begin{pmatrix} 0 & 1 & 0 \\ 1 & 0 & 0 \\ 0 & 0 & 0 \end{pmatrix}$} &,& \beta(k) \equiv \beta_{\text{o}} \scalebox{0.6}{$\begin{pmatrix} 0 & e^{-ik} & 0 \\ e^{ik} & 0 & 0 \\ 0 & 0 & 0 \end{pmatrix}$} , \nonumber \\
    \gamma(k) \equiv \gamma_{\text{o}} \scalebox{0.6}{$\begin{pmatrix} 0 & 1 & 0 \\ 1 & 0 & 0 \\ 0 & 0 & 1 \end{pmatrix}$} &,& 
\delta(k) \equiv \delta_{\text{o}} \scalebox{0.6}{$\begin{pmatrix} 1 & 0 & 0 \\ 0 & -1 & 0 \\ 0 & 0 & 1 \end{pmatrix}$} ,
\end{eqnarray}
where $\alpha_0$, $\beta_0$, $\gamma_0$, and $\delta_0$ represent the corresponding perturbation strengths. Under each of these perturbations, the real space Floquet operator respectively becomes
%%%%%%%%%%%%%%%%%%%%%%%

\begin{align}
    \hat{U}_{\alpha}  &= e^{(-i h_{2} T / 2)} e^{(-i h_{1,\alpha} T / 2)} , \label{eq:eqa} \\
    \hat{U}_{\beta}  &= e^{(-i h_{2,\beta} T / 2)} e^{(-i h_{1} T / 2)} , \label{eq:eqb} \\
    \hat{U}_{\gamma}  &= e^{(-i h_{2} T / 2)} e^{(-i h_{1,\gamma} T / 2)} , \label{eq:eqg} \\
    \hat{U}_{\delta}  &= e^{(-i h_{2,\delta} T / 2)} e^{(-i h_{1,\delta} T / 2)} , \label{eq:eqd}
\end{align}
where $h_1$ and $h_2$ are as previously defined,
\begin{eqnarray}
h_{1,\alpha} &=& \sum_{j=1}^{N}  \left(J_1\,S_x + S_\alpha \right) \otimes |j\rangle \langle j| , \nonumber \\    
h_{2,\beta} &=& \sum_{j=1}^{N-1}  \left(J_{2}\,S_+ + S_\beta \right) \otimes |j\rangle \langle j+1|   + h.c. , \nonumber \\
h_{1,\gamma} &=& \sum_{j=1}^{N}  \left( J_1\,S_x + S_\gamma \right) \otimes |j\rangle \langle j| , \nonumber \\
h_{1,\delta} &=& \sum_{j=1}^{N} \left( J_1 \, S_x + S_\delta \right) \otimes |j\rangle \langle j| , \nonumber \\
h_{2,\delta} &=& \sum_{j=1}^{N-1} \left[ J_{2} \, S_+ \otimes |j\rangle \langle j+1|  + h.c.\right] + \sum_{j=1}^{N} S_\delta \otimes |j\rangle \langle j| , \nonumber \\
\end{eqnarray}
and
\begin{eqnarray}
    S_\alpha &=& \begin{pmatrix} 
0 & \alpha_{\text{o}} & 0 \\ 
\alpha_{\text{o}} & 0 & 0 \\ 
0 & 0 & 0 
\end{pmatrix}, \nonumber \\
S_\beta &=& \begin{pmatrix} 
0 & \beta_{\text{o}} & 0 \\ 
0 & 0 & 0 \\ 
0 & 0 & 0 
\end{pmatrix}, \nonumber \\
S_\gamma &=& \begin{pmatrix} 
0 & \gamma_{\text{o}} & 0 \\ 
\gamma_{\text{o}} & 0 & 0 \\ 
0 & 0 & \gamma_{\text{o}} 
\end{pmatrix}, \nonumber \\
S_\delta &=& \begin{pmatrix} 
\delta_{\text{o}} & 0 & 0 \\ 
0 & -\delta_{\text{o}} & 0 \\ 
0 & 0 & \delta_{\text{o}} 
\end{pmatrix}.
\end{eqnarray}

The perturbations $\alpha(k)$ and $\beta(k)$ introduce an imbalance in the intracell and intercell hopping amplitudes, respectively. The perturbation $\gamma(k)$ is designed to create an on-site potential on sublattice $C$ while also creating an imbalance in the intracell hopping amplitudes. Meanwhile, the perturbation $\delta(k)$ introduces on-site potentials across all three sublattices. The impact of all the introduced perturbations on the system's symmetries is summarized in Table~\ref{tbl:tbl1}. In short, $\alpha(k)$ and $\beta(k)$ preserve all chiral, time-reversal, and particle-hole symmetries, whereas $\gamma(k)$ and $\delta(k)$ break some of these symmetries. 

\begin{table}[ht]
    \centering
    \resizebox{0.35\textwidth}{!}{
    \begin{tabular}{|>{\columncolor{gray!10}}c|c|c|c|c|}
        \hline
         \textbf{Symmetry} & $\alpha(k)$ & $\beta(k)$ & $\gamma(k)$ & $\delta(k)$ \\ 
        \hline
        Chiral & \cmark & \cmark & \xmark & \xmark \\ \hline
         Time-reversal & \cmark & \cmark & \cmark & \cmark \\ \hline
        Particle-hole & \cmark & \cmark & \xmark & \xmark \\ \hline
    \end{tabular}}
    \caption{A summary of the symmetries preserved and broken by each perturbation operator discussed in Sec.~\ref{pert}.}
    \label{tbl:tbl1}
\end{table}

\begin{center}
  \begin{figure}[htpb]
  \includegraphics[width=0.45\textwidth]{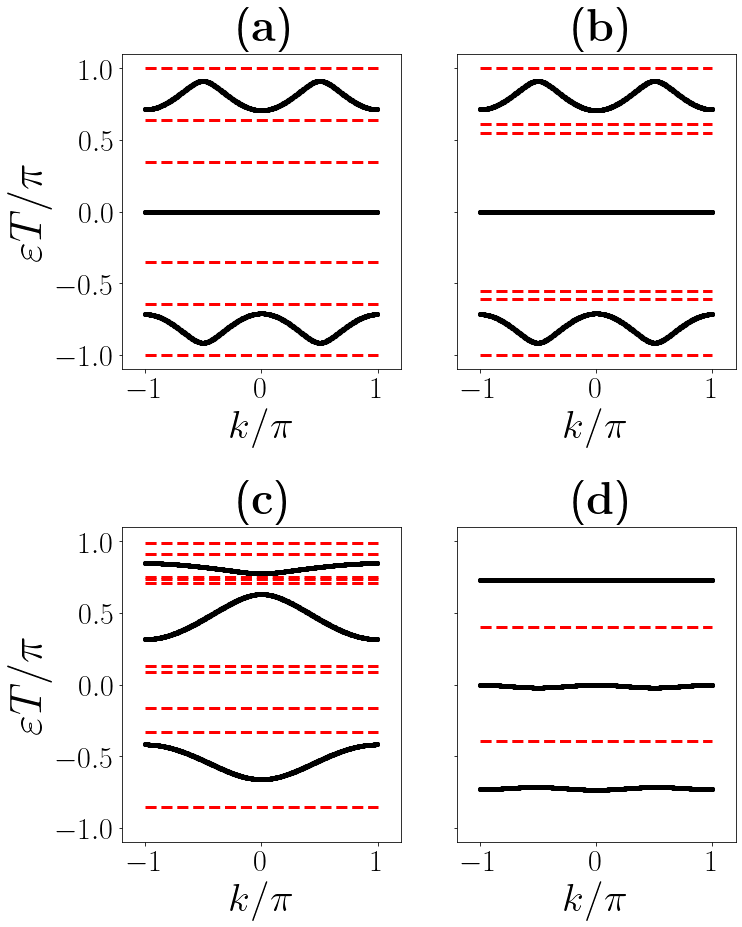}
  \captionof{figure}{The momentum space quasienergy spectrum in the presence of each of the four perturbations $\alpha(k)$, $\beta(k)$, $\gamma(k)$, and $\delta(k)$. Subplot (a) is at $J_{1}= 1.4 \pi$, $J_{2}= 1.2\pi$, and $\alpha_{o} = 1.25\pi$, subplot (b) is at  $J_{1}= 1.2 \pi$, $J_{2}= 1.4\pi$, and $\beta_{o} = 1.25\pi$, subplot (c) is at  $J_{1}= 0.2 \pi$, $J_{2}= 1.3\pi$, and $\gamma_{o} = 1.25\pi$, and subplot (d) is at  $J_{1}= 0.2 \pi$, $J_{2}= 1.3\pi$, and $\delta_{o} = 2\pi$. The red dashed lines mark the quasienergies of the expected edge states when OBC are applied (see Fig.~\ref{fig:fig5})}
  \label{fig:fig4}
\end{figure} 
\end{center}

\begin{center}
  \begin{figure}[htpb]
  \includegraphics[width=0.45\textwidth]{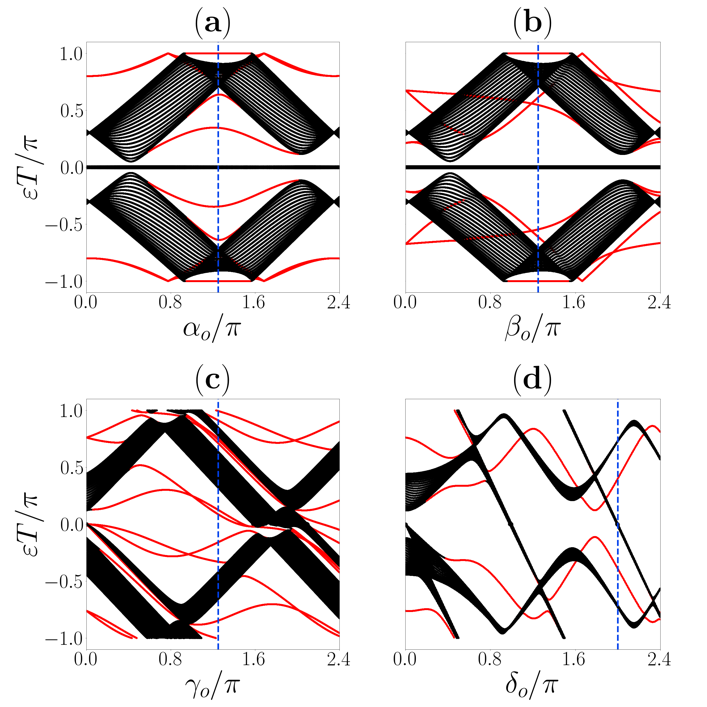}
  \captionof{figure}{The quasienergy spectrum versus the perturbation strengths for Eqs.~(\ref{eq:eqa})-(\ref{eq:eqd}) in a system of $N=25$ unit cells at $T=2$. (a) is at $J_{1}= 1.4 \pi$ and $J_{2}= 1.2\pi$, (b) is at  $J_{1}= 1.2 \pi$ and $J_{2}= 1.4\pi$, (c) is at  $J_{1}= 0.2 \pi$ and $J_{2}= 1.3\pi$, and subplot (d) is at  $J_{1}= 0.2 \pi$ and $J_{2}= 1.3\pi$. The edge modes are represented in red. The blue vertical dashed lines mark the parameter values used in Fig.~\ref{fig:fig4}.}
  \label{fig:fig5}
\end{figure}    
\end{center}

To further reveal the system's behavior under the influence of the aforementioned perturbations, in Figs.~\ref{fig:fig4} and~\ref{fig:fig5} we show the energy spectra of our model in momentum space and real space, respectively. By comparing the two figures, and noticing that Fig.~\ref{fig:fig4} depicts just the bulk states in the momentum space energy spectrum, we can easily distinguish between bulk and edge states in the real space energy spectrum given by Fig.~\ref{fig:fig5}.

In Fig.~\ref{fig:fig5}(a), the system is only subject to perturbation $\alpha(k)$. Upon comparing with Fig.~\ref{fig:fig2}(a), we observe that the edge states that exist in the absence of any perturbation are preserved at small $\alpha_0$. At large enough $\alpha_0$, a gap closes and reopens at quasienergy $0$, followed by the emergence of additional edge states within the two bulk bands. When $\alpha_0$ further increases, another gap closing and reopening occurs at quasienergy $\pi/T$. Remarkably, this gap-closing and reopening event is followed by the emergence of the so-called $\pi$ modes (edge states that are pinned at quasienergy $\pi/T$). The presence of such $\pi$ modes is a consequence of the interplay between chiral symmetry and topology. 

Specifically, as chiral symmetry forces Floquet eigenstates to come in pairs with quasienergy $ \varepsilon$, $\pm\varepsilon=0, \pi/T$ are the only values of quasienergies whose corresponding eigenstates can simultaneously be eigenstates of the chiral symmetry operator. As a result, zero and $\pi$ edge modes are special as they are protected by the discrete nature of chiral symmetry eigenstates. In a three-band chiral symmetric model, one bulk band necessarily occupies zero energy, making it impossible for zero edge modes to exist. As $\pi$ edge modes are unique to Floquet systems, a static three-band chiral symmetric system cannot host a chiral symmetry-protected edge state and instead only supports pairs of edge states at positive and negative energy values~\cite{Ghuneim2024, Du2024, Alvarez2019}. On the other hand, our result above not only explicitly demonstrates the emergence of a chiral symmetry-protected edge state in our (time-periodic) system, but also its coexistence with ordinary positive and negative energy edge state pairs found in its static counterpart.

Apart from adding the perturbation $\alpha$, the above $\pi$ edge modes could also arise in the presence of perturbation $\beta$, as depicted in Fig.~\ref{fig:fig5}(b). Apart from this similarity, the perturbations $\alpha$ and $\beta$ have different effects on the system's existing edge states. In particular, while the perturbation $\alpha$ yields a significant quasienergy splitting between the left- and right-localized edge states within the two bulk bands (edge states closest to zero quasienergy), the perturbation $\beta$ instead results in a significant splitting between the edge states beyond the two bulk bands (edge states closest to $\pi/T$ quasienergy). This different effect could be understood by first noting that $\alpha$ is intrasite in nature, whereas the perturbation $\beta$ is intersite. In this case, $\alpha$ will have a significant effect on quasienergy eigenstates that have significant peaks on sites within the unit cell, e.g., the edge states of Fig.~\ref{fig:fig3}(b). On the other hand, $\beta$ instead has a significant effect on quasienergy eigenstates with significant peaks on sites from adjacent unit cells, e.g., the edge states of Fig.~\ref{fig:fig3}(a).

As the other two perturbations $\gamma$ and $\delta$ break chiral symmetry, the resulting quasienergy structure (including both bulk and edge states) is not symmetrical about $\varepsilon=0$. Moreover, while such perturbations also open the gap at $\pi/T$ quasienergy, they do not lead to the emergence of $\pi$ edge modes. This is consistent with the fact that $\pi$ edge modes are protected by the chiral symmetry. 

In Appendix~\ref{app:B}, we presented another set of quasienergy plots with respect to each perturbation above at different parameter values and at a larger range of perturbation strengths. There, we not only find the emergence of $\pi$ edge modes for the cases of $\alpha$ and $\beta$ perturbations over a window of parameter values (suggesting that the presence of $\pi$ edge modes is not fine-tuned to specific values of $\alpha_o$ and $\beta_o$), but also verify that the quasienergy structure qualitatively repeats itself as the perturbation strength is increased.

%%%%%%%%%%%%

\subsection{Spatial disorder effects}
\label{dis}
\begin{center}
    \begin{figure}[htpb]
    \begin{subfigure}[l]{0.22\textwidth}
        \includegraphics[width=\linewidth]{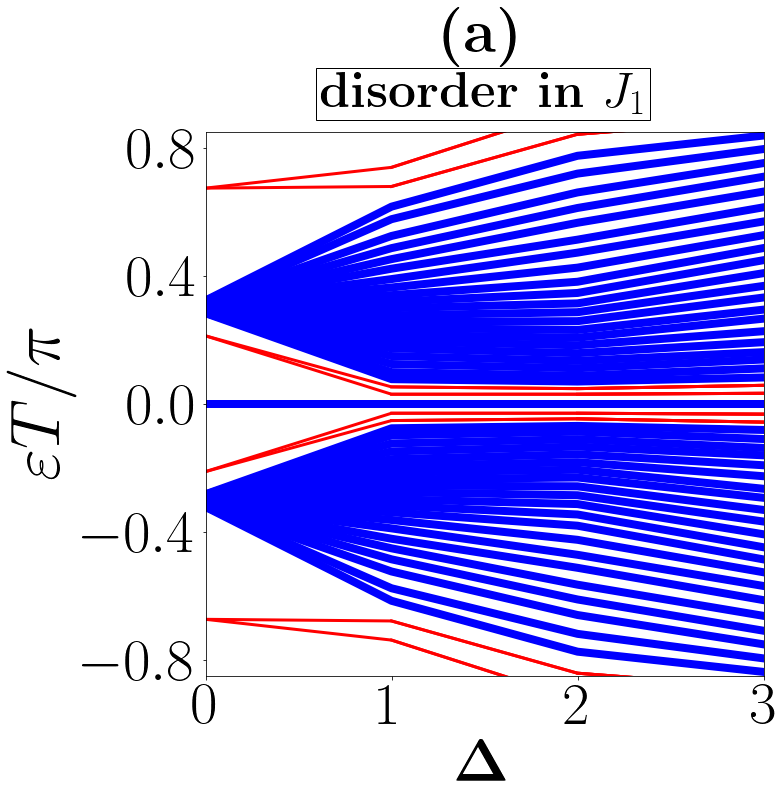}   
    \end{subfigure}%
    \begin{subfigure}[r]{0.22\textwidth}
        \includegraphics[width=\linewidth]{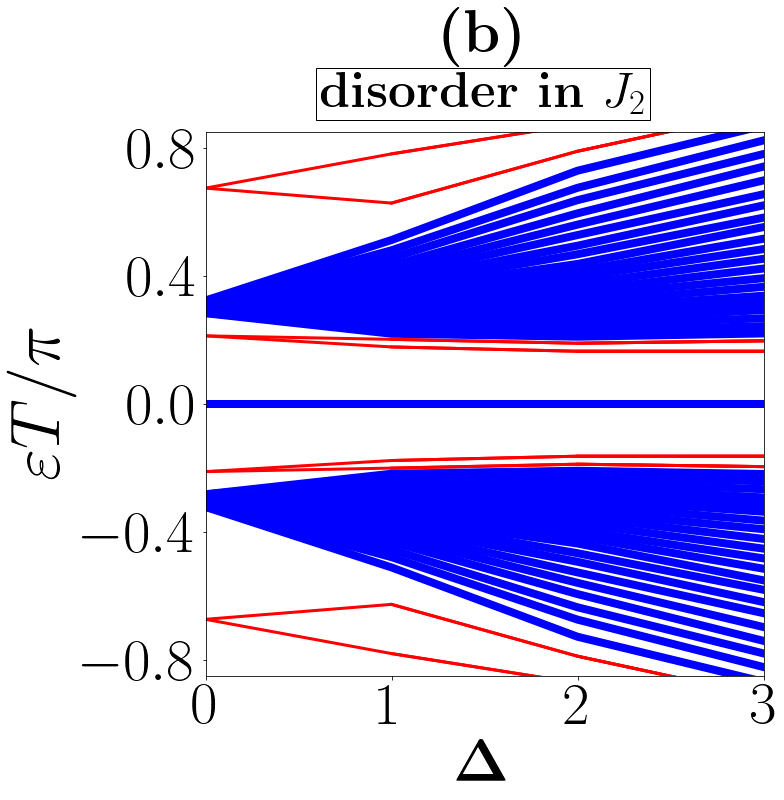}     
    \end{subfigure}%
    \vspace{0.2cm}  
    \begin{subfigure}[r]{0.22\textwidth}
        \includegraphics[width=\linewidth]{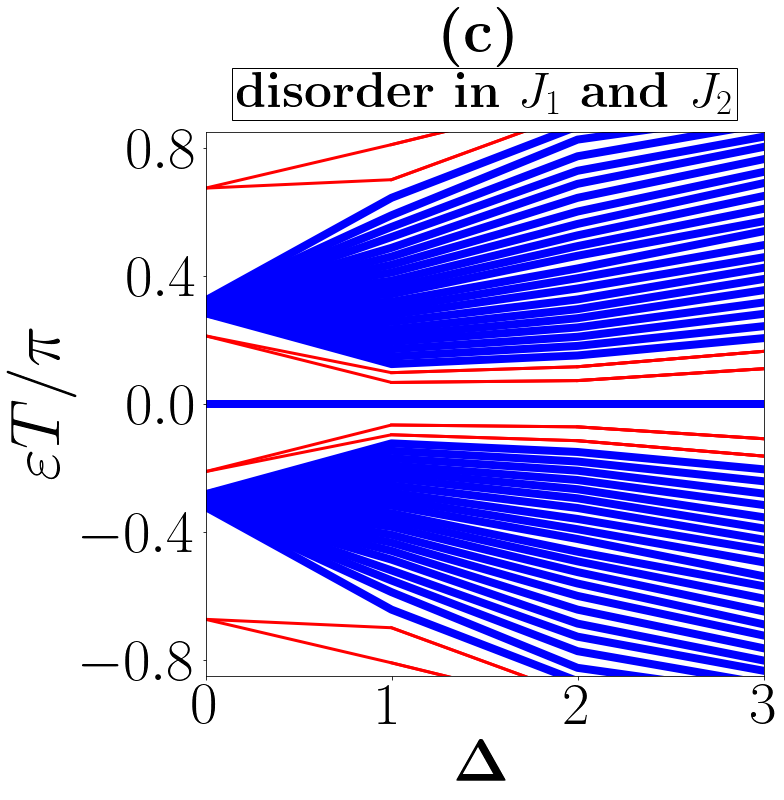}   
    \end{subfigure}%  
    \caption{The quasienergy spectrum of disordered Eq.~(\ref{eq:eq3}) versus the disorder strength for $N = 25$ unit cells under OBC. The starting parameter values for all three cases are $J_{1} = 1.2\pi$ and $J_{2} = 1.4\pi$. (a) Disorder acts only on $J_{1}$, (b) Disorder acts only on $J_{2}$, and (c) Disorder acts on both $J_{1}$ and $J_{2}$. The edge modes are represented in red. Each data point is averaged over 100 disorder realizations.}
    \label{fig:fig6}
\end{figure}
\end{center}

The spatial disorder is a crucial aspect when assessing the stability and behavior of quantum systems, particularly those having topological phases. In realistic contexts, flaws such as random fluctuations in system parameters are unavoidable, and adding spatial disorder into theoretical models helps to bridge the gap between idealized and real-world systems. In this section, we inject disorder into our time-periodic system via the parameter $\iota_{r,j}$ that modify each coupling parameter $J_{r}$ ($r=1,2$) such that $J_{r} \rightarrow J_{r} + \iota_{r,j}$, where $j=1,2,\cdots, N$ and $\iota_{r,j}$ is taken randomly from [$-\Delta$, $\Delta$]. The parameter $\Delta$ is referred to as the disorder strength. For completeness, we begin by introducing disorder separately to the intracell and intercell hopping parameters to analyze the individual effect on each parameter. We then introduce disorder to both hopping parameters simultaneously to investigate their combined impact on the system's edge states and overall robustness, which will be especially relevant in the experimental setting where spatial disorder is present for both hopping parameters.
%%%%%%%%%%%%%%%%%
Figure~\ref{fig:fig6} summarizes our findings, demonstrating the overall robustness of the system's edge states against disorder. Indeed, while each pair of edge states is split in quasienergy, they are still present under all values of disorder strength we considered. Moreover, we further observe that most of these edge states seem to approach but do not actually touch a bulk band. On the other hand, the two outer edge states remain well separated from any bulk band.

\begin{center}
    \begin{figure}[ht]
        \begin{subfigure}[t]{0.24\textwidth}
        \includegraphics[width=\linewidth]{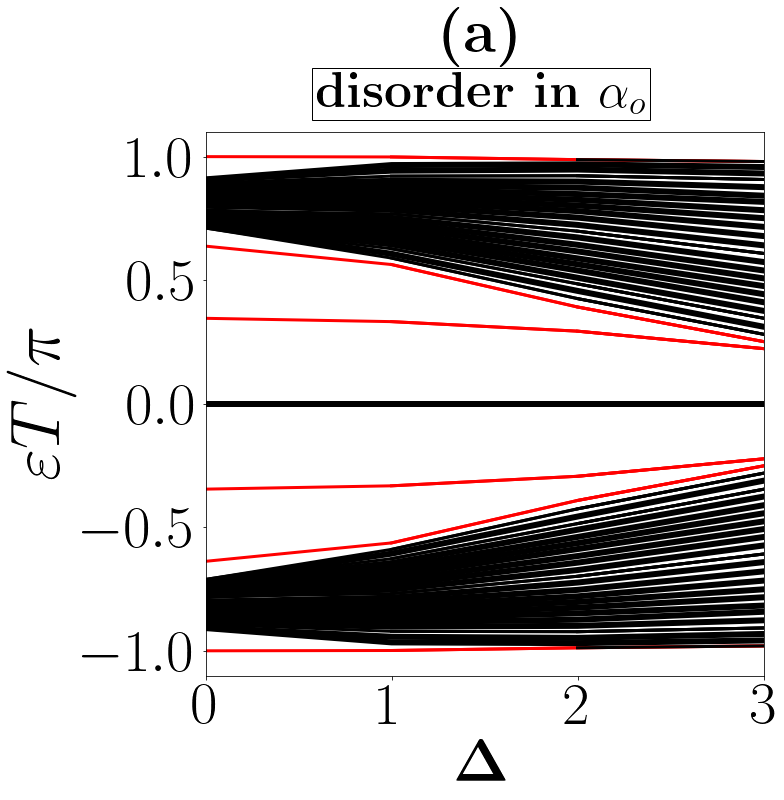}   
        \end{subfigure}%
        \hspace{0.2cm}
        \begin{subfigure}[t]{0.2173\textwidth}
        \includegraphics[width=\linewidth]{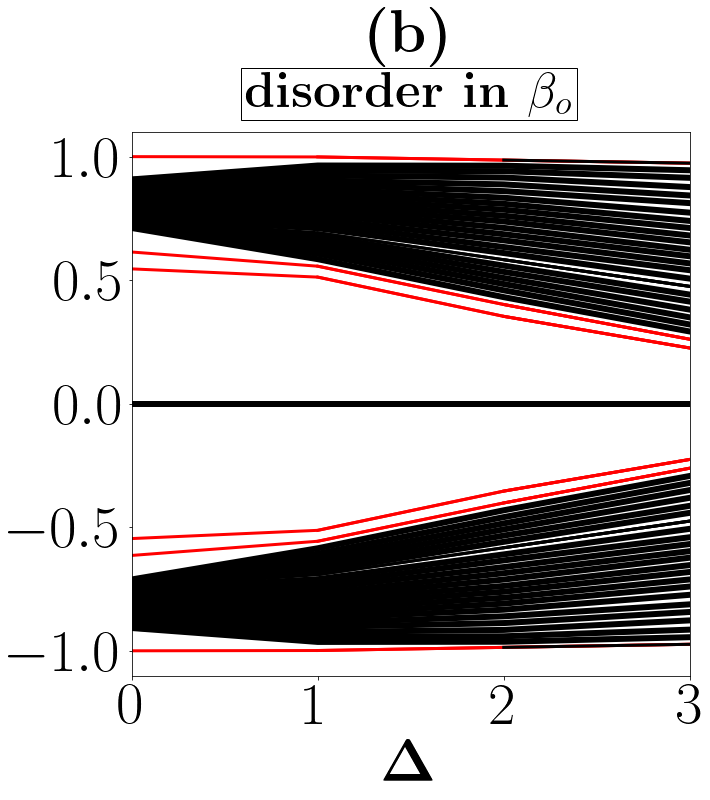}     
        \end{subfigure}%      
        \vspace{0.25cm} 
        \begin{subfigure}[t]{0.24\textwidth}
        \includegraphics[width=\linewidth]{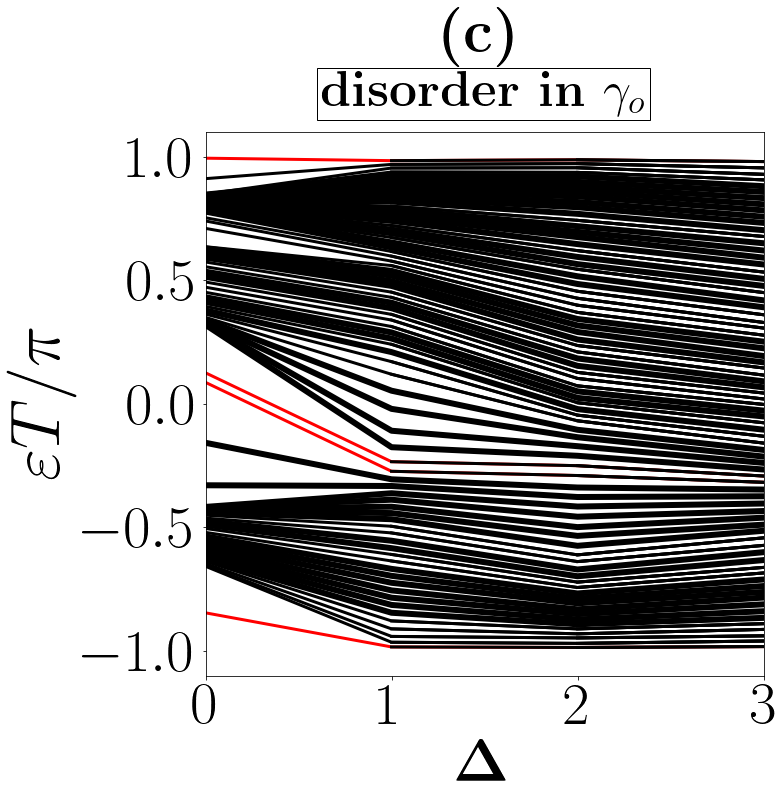}   
        \end{subfigure}% 
        \hspace{0.2cm}
        \begin{subfigure}[t]{0.2173\textwidth}
        \includegraphics[width=\linewidth]{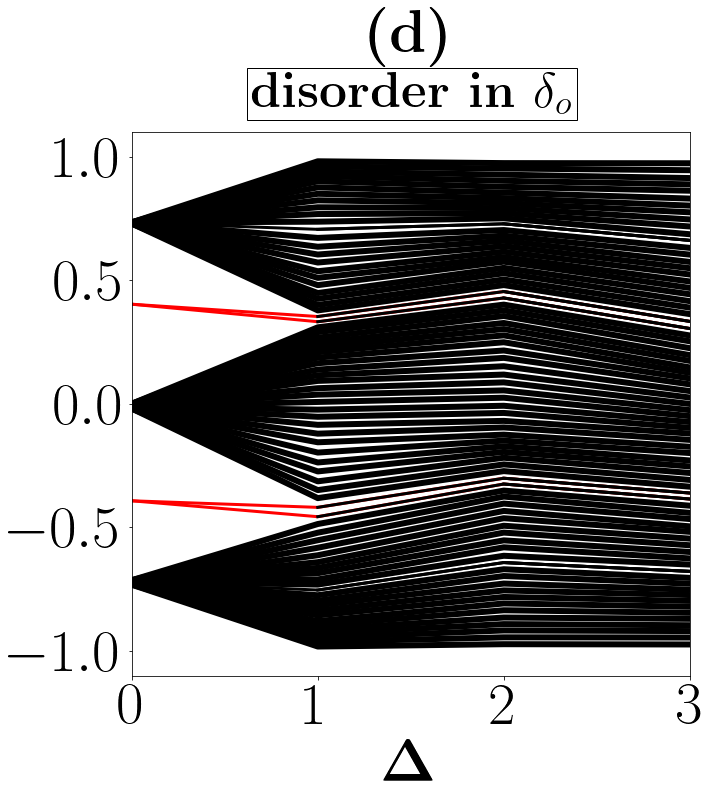}   
        \end{subfigure}%   
    \caption{The quasienergy spectrum versus the perturbation disorder strength of disordered Eqs.~(\ref{eq:eqa})-(\ref{eq:eqd}) for $N = 25$ unit cells under OBC. (a) includes only the disorder in $\alpha_{o}$ at the starting point of $\alpha_{o} = 1.25\pi$ and fixed $J_{1} = 1.4\pi$ and $J_{2} = 1.2\pi$, (b) includes only the disorder in $\beta_{o}$ at the starting point of $\beta_{o} = 1.25\pi$ and fixed $J_{1} = 1.4\pi$ and $J_{2} = 1.2\pi$, (c) includes only the disorder in $\gamma_{o}$ at the starting point of $\gamma_{o} = 1.25\pi$ and fixed $J_{1} = 0.2\pi$ and $J_{2} = 1.3\pi$, and (d) includes only the disorder in $\delta_{o}$ at the starting point of $\delta_{o} = 2\pi$ and fixed $J_{1} = 0.2\pi$ and $J_{2} = 1.3\pi$. The edge modes are represented in red. Each data point is averaged over 100 disorder realizations.}
    \label{fig:fig7}
\end{figure}
\end{center}

%%%%%%%%%%%%%%%%
In Fig.~\ref{fig:fig7}, we show the fate of the edge states in the simultaneous presence of each of the previously introduced perturbations ($\alpha, \beta, \gamma,$ and $\delta$) and spatial disorder in the respective perturbation parameter. In all cases, we find that the system's existing edge modes, including the previously obtained $\pi$-edge modes for the cases of $\alpha$ and $\beta$ perturbations, remain plainly visible and distinctly separated from their closest bulk band at least up to moderate disorder strength, demonstrating their robustness and stability against both perturbation and disorder simultaneously.

%%%%%%%%%%%%%%%%%%%

\section{Conclusion}
\label{conc}
In this work, we investigated the properties of a periodically driven extended SSH model that consists of three sites per unit cell, paying particular attention to the formation and behavior of edge states. We further identified the presence of chiral, particle-hole and time-reversal symmetries in the system. To uncover the role of these symmetries on the system's edge states, we considered the effect of four representative perturbations separately, two of which preserve all symmetries while the remaining two break some symmetries. Interestingly, under the two perturbations that preserve all symmetries, not only did we confirm the robustness of the existing edge states, but we also observed the opening of an additional gap at $\pi/T$ quasienergy that is followed by the emergence of $\pi$ edge modes, a hallmark of Floquet symmetry-protected topological systems. Under the other two perturbations that break some symmetries, the edge states remain present, though their structure becomes irregular. Finally, we have considered the effect of spatial disorder on each hopping parameter and each of the previously introduced perturbations, then confirmed that the system's edge states are resilient against these various disorders. 

In the future, it would be natural to extend the present model further to involve a larger number of sites per unit cell. This amounts to replacing $S_x$ and $S_y$ in our Hamiltonian description by their $n\times n$ variants with $n>3$. In addition to the potential novel edge state structures such a generalized SSH model may offer, an important motivation for studying such a model is to investigate the extent to which the above obtained $\pi$ edge modes could emerge. While it is reasonable to believe that such $\pi$ edge modes are easily obtained in models with an even number of bands ($n$ is even), it is not immediately clear if $\pi$ edge modes also arise in models with an odd number of bands (apart from the case of $n=3$ presently considered in the paper). 

Other interesting potential directions to pursue include the incorporation of interaction \cite{Koor2022, Yates2019, von2016}, nonlinearity \cite{Chaunsali2021, Tuloup2020}, and/or non-Hermiticity \cite{Shen2024, Vyas2021, Wu2020, Shi2024} on the present model. One could further consider a higher-dimensional extension of the present model to yield novel higher-order topological phases that host interesting corner and/or hinge modes. Finally, a more sophisticated periodic-driving scheme could be devised instead of the simplest two-step drive considered in the present paper to yield richer edge state structures, such as the emergence of multiple $\pi$ edge modes in the same edge \cite{Bomantara2022, Zhou2022, Zhou2020, Pan2020, Longwen2022, Wu2023}.

\begin{center}
\textbf{Data Availability}
\end{center}

The data that support the findings of this article are openly available \cite{Zenodo2025}.

 \begin{acknowledgements}
 This work was supported by the Deanship of Research
Oversight and Coordination (DROC) at King Fahd University of Petroleum \& Minerals (KFUPM) through Project No.~EC221010, as well as by the Deanship of Research Oversight and Coordination (DROC) and the Interdisciplinary Research Center(IRC) for Intelligent Secure Systems (ISS) at King Fahd University of Petroleum \& Minerals (KFUPM) through internal research Grant No. INSS2507.
 \end{acknowledgements}

\clearpage
\appendix

%\section{Appendix section} 
%\label{app:A}

%Put lengthy math calculations and additional plots here....

\section{Additional numerical results} 
\label{app:B}
\begin{figure}[htpb]
    \centering
    \begin{subfigure}[b]{0.4\textwidth}
        \includegraphics[width=\textwidth]{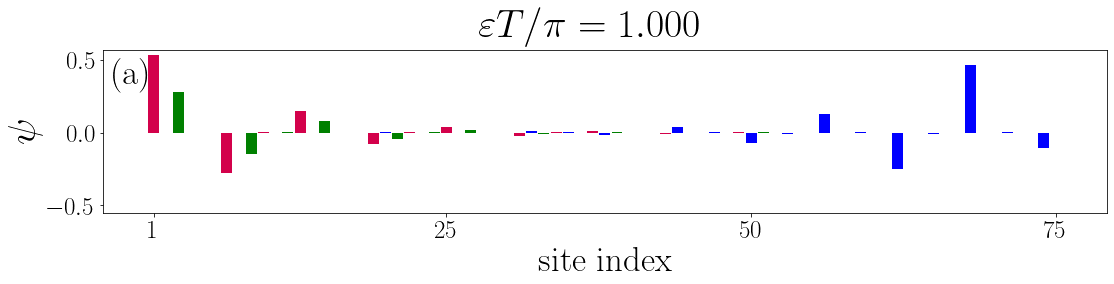} 
        \label{fig:figa1}
    \end{subfigure}
    \vspace{-0.5em} % Adjust this value as needed
    \begin{subfigure}[b]{0.4\textwidth}
        \includegraphics[width=\textwidth]{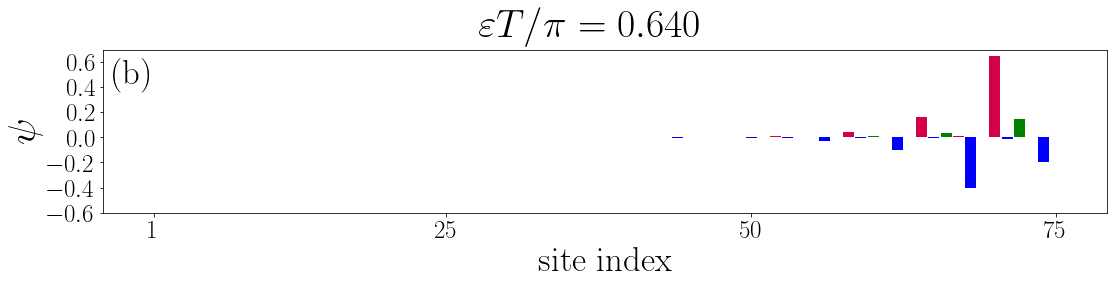}
        \label{fig:figa2}
    \end{subfigure}
    \vspace{-0.5em}
    \begin{subfigure}[b]{0.4\textwidth}
        \includegraphics[width=\textwidth]{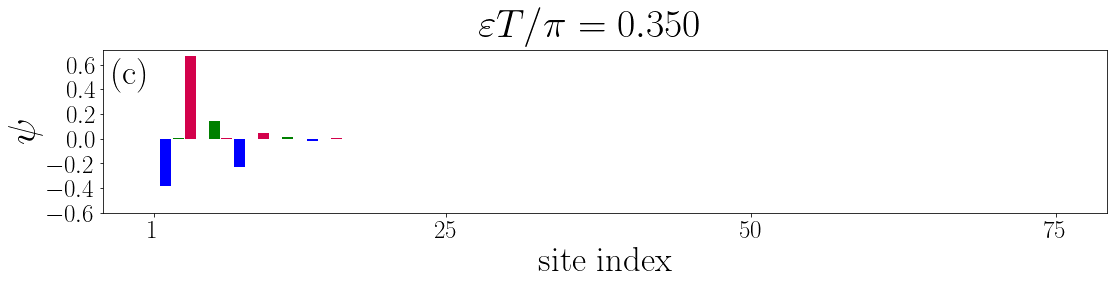} 
        \label{fig:figa3}
    \end{subfigure}
    \vspace{-0.5em}
    \begin{subfigure}[b]{0.4\textwidth}
        \includegraphics[width=\textwidth]{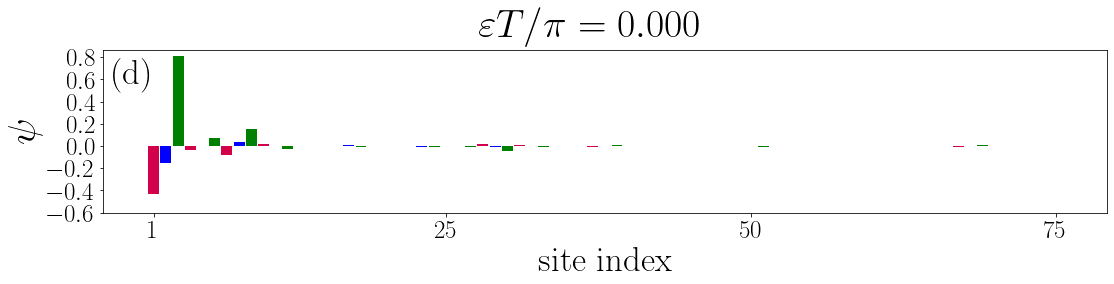}
        \label{fig:figa4}
    \end{subfigure}
    \vspace{-0.5em}
    \begin{subfigure}[b]{0.4\textwidth}
        \includegraphics[width=\textwidth]{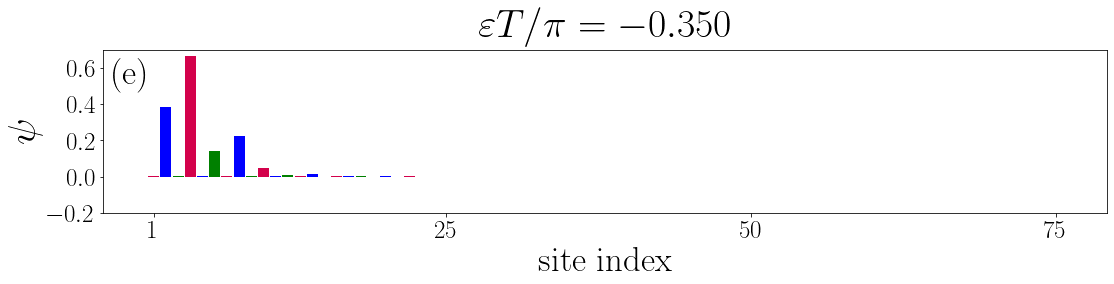}
        \label{fig:figa5}
    \end{subfigure}
    \vspace{-0.5em}
    \begin{subfigure}[b]{0.4\textwidth}   
        \includegraphics[width=\textwidth]{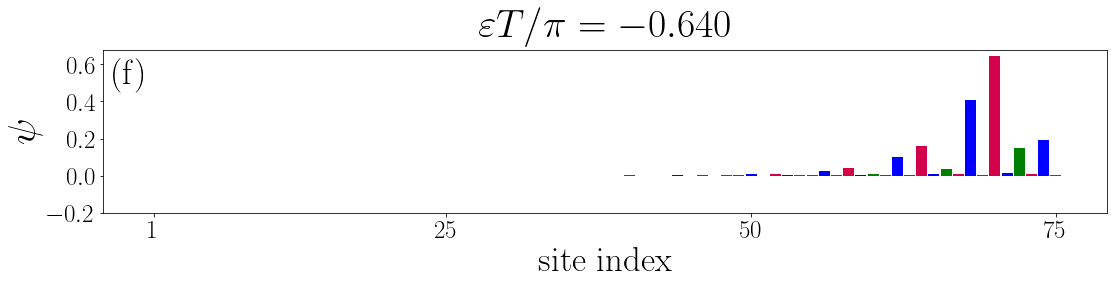} 
        \label{fig:figa6}
    \end{subfigure}
    %\begin{subfigure}[b]{0.4\textwidth}
        %\includegraphics[width=\textwidth]{a7}        
        %\label{fig:figa7}
    %\end{subfigure}
    \caption{Wave function profiles associated with the edge states that exist in the presence of $\alpha$ perturbation. The system parameters are fixed at $J_{1}= 1.4\pi$, $J_{2}=1.2\pi$, and $\alpha_{o} = 1.25\pi$. Note that a pair of edge state profiles (one left-localized and one right-localized) are plotted together for the case of $\varepsilon T/\pi = 1$.}
    \label{fig:fig8}
\end{figure}

\begin{figure}[htpb]
    \centering
    \begin{subfigure}[b]{0.4\textwidth}
        \includegraphics[width=\textwidth]{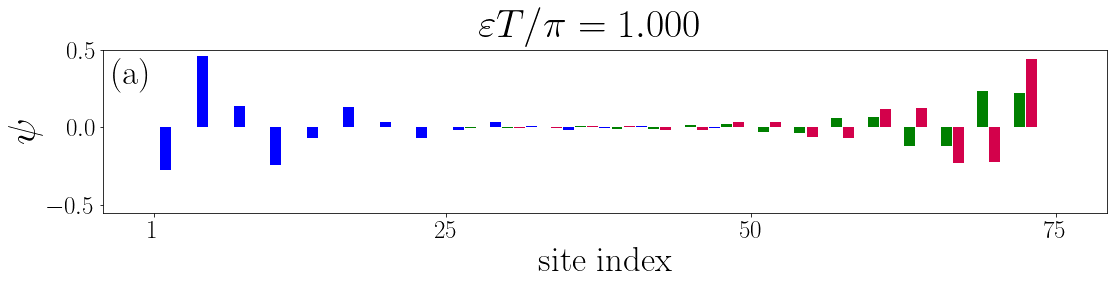} 
        \label{fig:figb1}
    \end{subfigure}
    \vspace{-0.5em} % Adjust this value as needed
    \begin{subfigure}[b]{0.4\textwidth}
        \includegraphics[width=\textwidth]{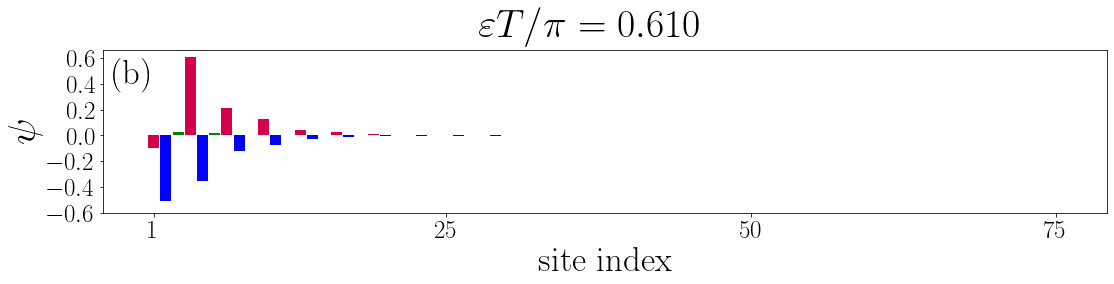}
        \label{fig:figb2}
    \end{subfigure}
    \vspace{-0.5em}
    \begin{subfigure}[b]{0.4\textwidth}
        \includegraphics[width=\textwidth]{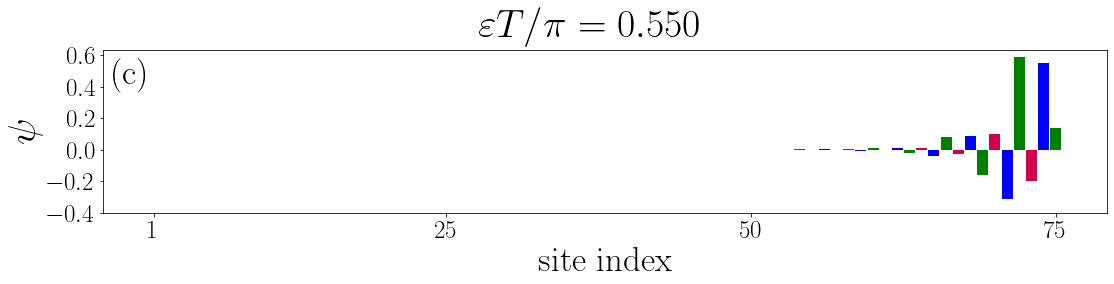} 
        \label{fig:figb3}
    \end{subfigure}
    \vspace{-0.5em}
    \begin{subfigure}[b]{0.4\textwidth}
        \includegraphics[width=\textwidth]{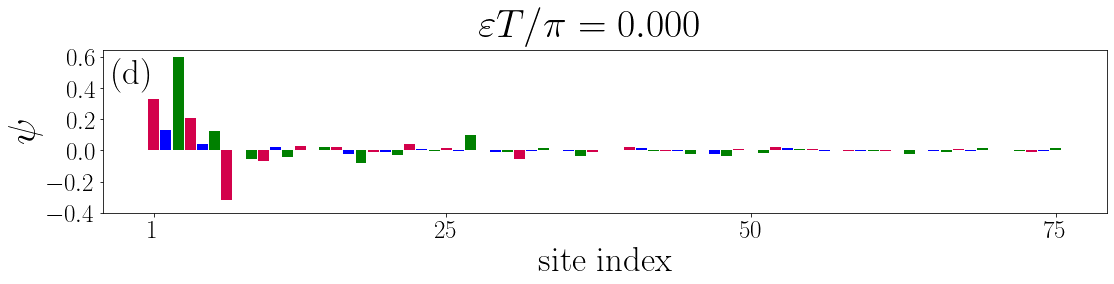}
        \label{fig:figb4}
    \end{subfigure}
    \vspace{-0.5em}
    \begin{subfigure}[b]{0.4\textwidth}
        \includegraphics[width=\textwidth]{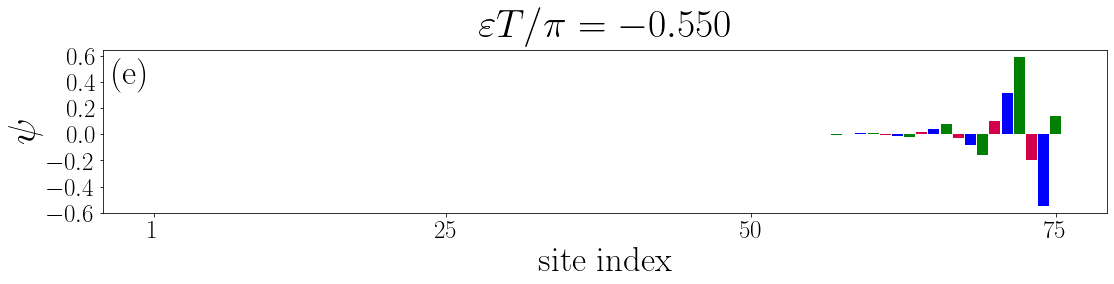}
        \label{fig:figb5}
    \end{subfigure}
    \vspace{-0.5em}
    \begin{subfigure}[b]{0.4\textwidth}   
        \includegraphics[width=\textwidth]{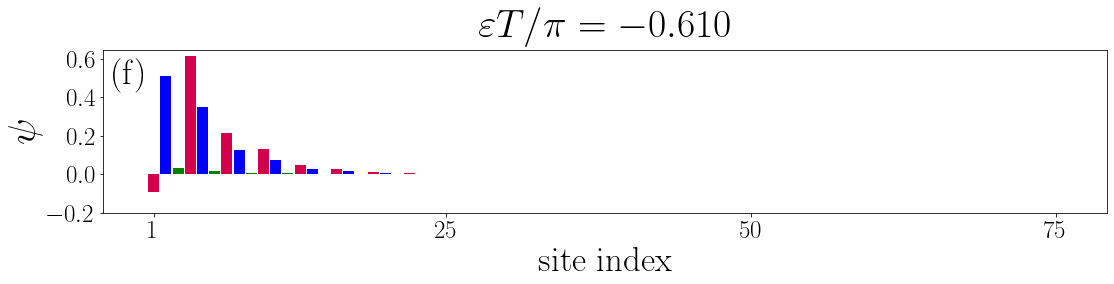} 
        \label{fig:figb6}
    \end{subfigure}
    %\begin{subfigure}[b]{0.4\textwidth}
        %\includegraphics[width=\textwidth]{b7}        
        %\label{fig:figb7}
    %\end{subfigure}
    \caption{Wave function profiles associated with the edge states that exist in the presence of $\beta$ perturbation. The system parameters are fixed at $J_{1}= 1.2\pi$, $J_{2}=1.4\pi$, and $\beta_{o} = 1.25\pi$. Note that a pair of edge state profiles (one left-localized and one right-localized) are plotted together for the case of $\varepsilon T/\pi = 1$.}
    \label{fig:fig9}
\end{figure}

\begin{figure}[htpb]
    \centering
    \begin{subfigure}[g]{0.4\textwidth}
        \includegraphics[width=\textwidth]{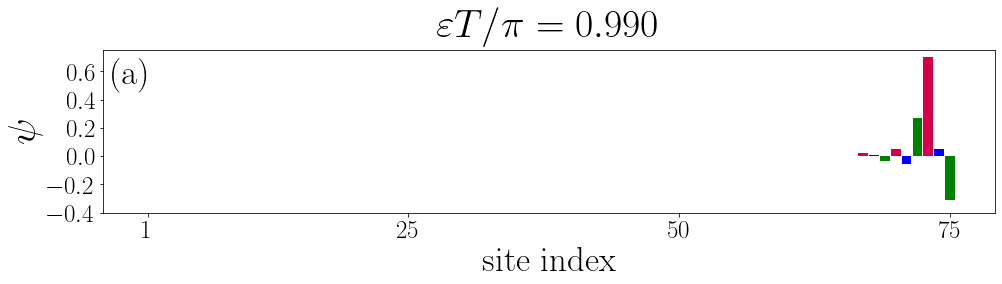} 
        \label{fig:figg1}
    \end{subfigure}
    \vspace{-0.5em} % Adjust this value to decrease/increase the space
    \begin{subfigure}[g]{0.4\textwidth}
        \includegraphics[width=\textwidth]{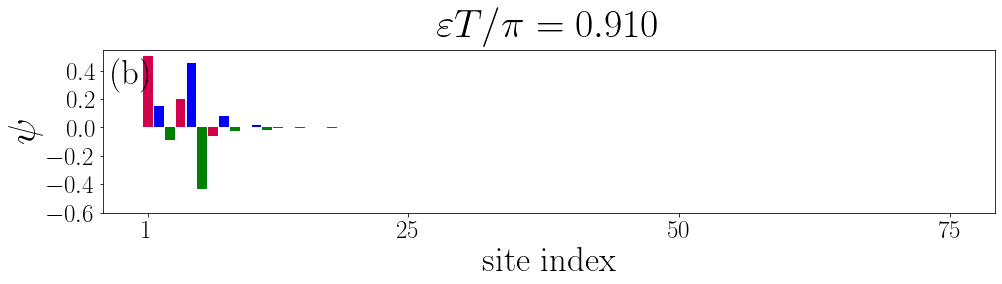}
        \label{fig:figg2}
    \end{subfigure}
    \vspace{-0.5em}
    \begin{subfigure}[g]{0.4\textwidth}
        \includegraphics[width=\textwidth]{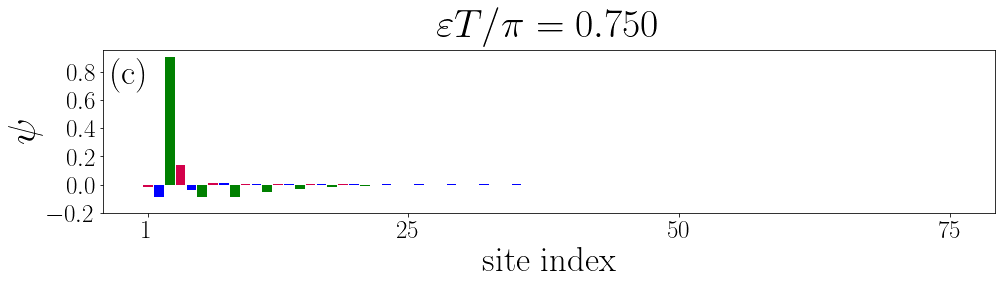} 
        \label{fig:figg3}
    \end{subfigure}
    \vspace{-0.5em}
    \begin{subfigure}[g]{0.4\textwidth}
        \includegraphics[width=\textwidth]{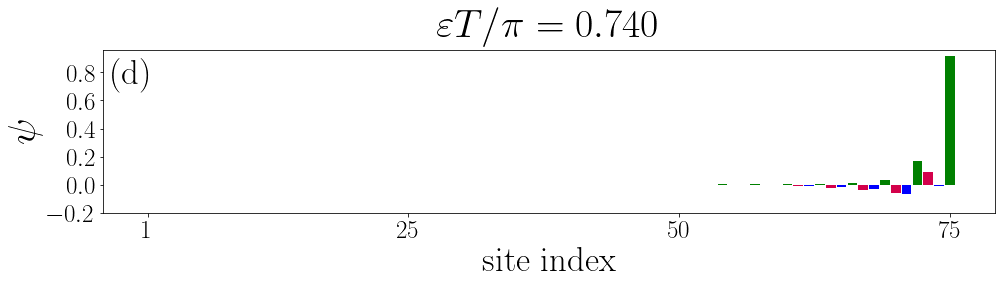}
        \label{fig:figg4}
    \end{subfigure}
    \vspace{-0.5em}
    \begin{subfigure}[g]{0.4\textwidth}
        \includegraphics[width=\textwidth]{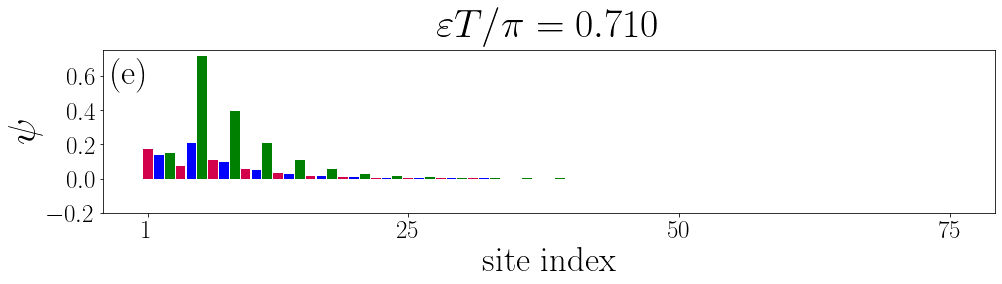}
        \label{fig:figg5}
    \end{subfigure}
    \vspace{-0.5em}
    \begin{subfigure}[g]{0.4\textwidth}
        \includegraphics[width=\textwidth]{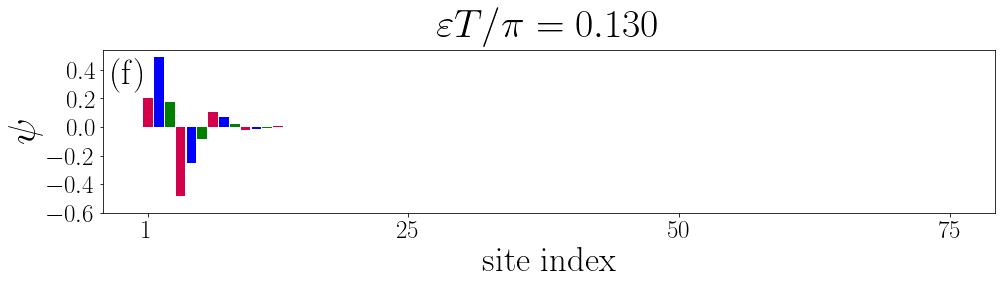} 
        \label{fig:figg6}
    \end{subfigure}
    \vspace{-0.5em}
    \begin{subfigure}[g]{0.4\textwidth}
        \includegraphics[width=\textwidth]{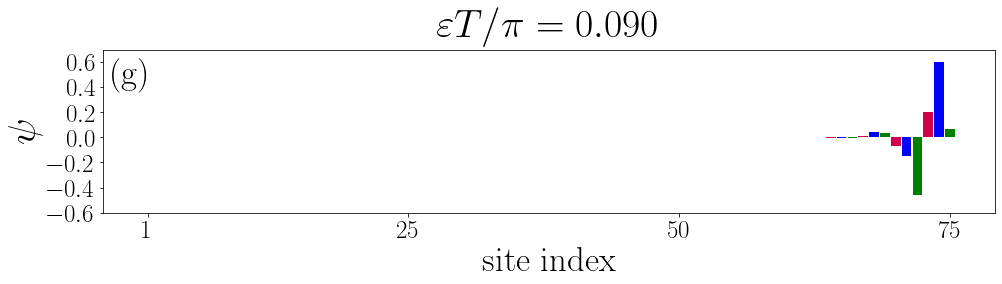}        
        \label{fig:figg7}
    \end{subfigure}
    \vspace{-0.5em}
    \begin{subfigure}[g]{0.4\textwidth}
        \includegraphics[width=\textwidth]{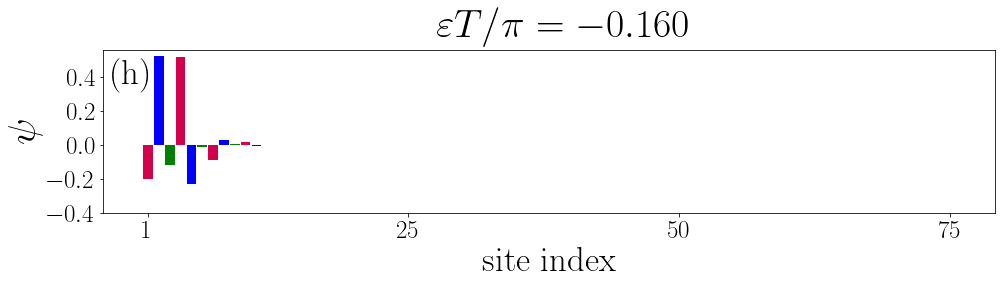}
        \label{fig:figg8}
    \end{subfigure}
    \vspace{-0.5em}
    \begin{subfigure}[g]{0.4\textwidth}
        \includegraphics[width=\textwidth]{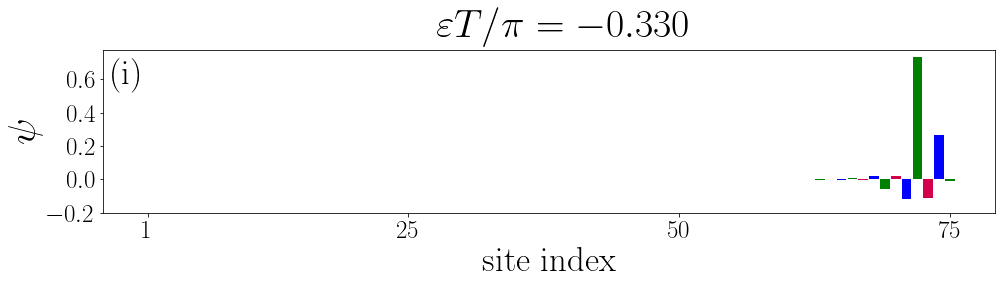}
        \label{fig:figg9}
    \end{subfigure}
    \vspace{-0.5em}
    \begin{subfigure}[g]{0.4\textwidth}
        \includegraphics[width=\textwidth]{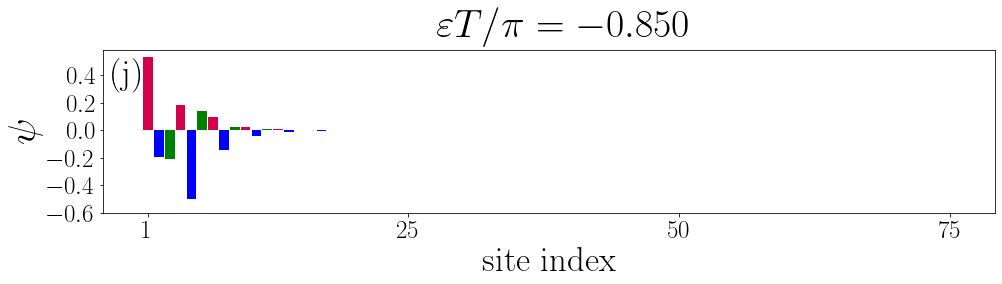}
        \label{fig:figg10}
    \end{subfigure}
\caption{Wave function profiles associated with the edge states that exist in the presence of $\gamma$ perturbation. The system parameters are fixed at  $J_{1}= 0.2\pi$, $J_{2}= 1.3\pi$, and $\gamma_{o} = 1.25\pi$.}
    \label{fig:fig10}
\end{figure}

\begin{figure}[htpb]
    \centering
    \begin{subfigure}[b]{0.4\textwidth}
        \includegraphics[width=\textwidth]{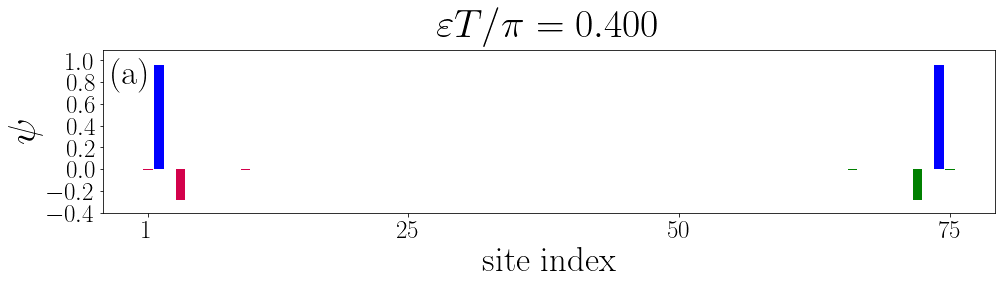} 
        \label{fig:figd1}
    \end{subfigure}
    \vspace{-0.5em}
    \begin{subfigure}[b]{0.4\textwidth}
        \includegraphics[width=\textwidth]{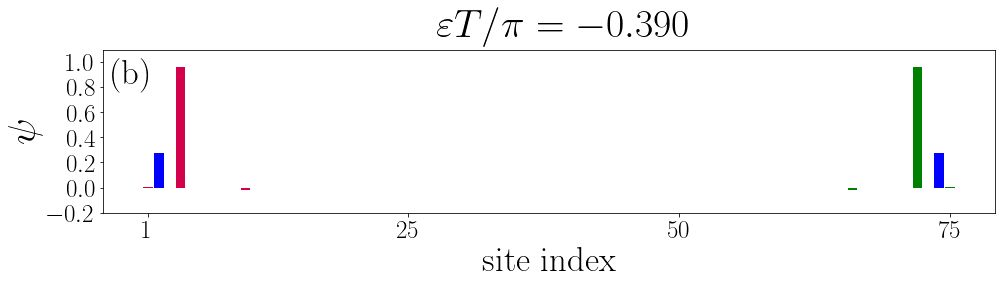}
        \label{fig:figd3}
    \end{subfigure}
    \caption{Wave function profiles associated with the edge states that exist in the presence of $\delta$ perturbation. The system parameters are fixed at $J_{1}= 0.2\pi$, $J_{2}=1.3\pi$, and $\delta_{o} = 2\pi$. Note that a pair of edge state profiles (one left-localized and one right-localized) are plotted together for each panel.}
    \label{fig:fig11}
\end{figure}

%%%%%%%%%%%%%%%%%%
In the main discussion, we demonstrated that the system's edge states remain robust under some representative perturbations. For completeness, this section displays the wave function profiles of the various edge states that exist under each of the four perturbations introduced in the main text. These are presented in Figs.~(\ref{fig:fig8})-(\ref{fig:fig11}).

%%%%%%%%%%%%%%%%%%%
    
%%%%%%%%%%%%%%%%%%%

%%%%%%%%%%%%%%
%%%%%%%%%%%%%%

%%%%%%%%%%%%%%%%%%%%

\begin{center}
  \begin{figure}[htpb]
  \includegraphics[width=0.45\textwidth]{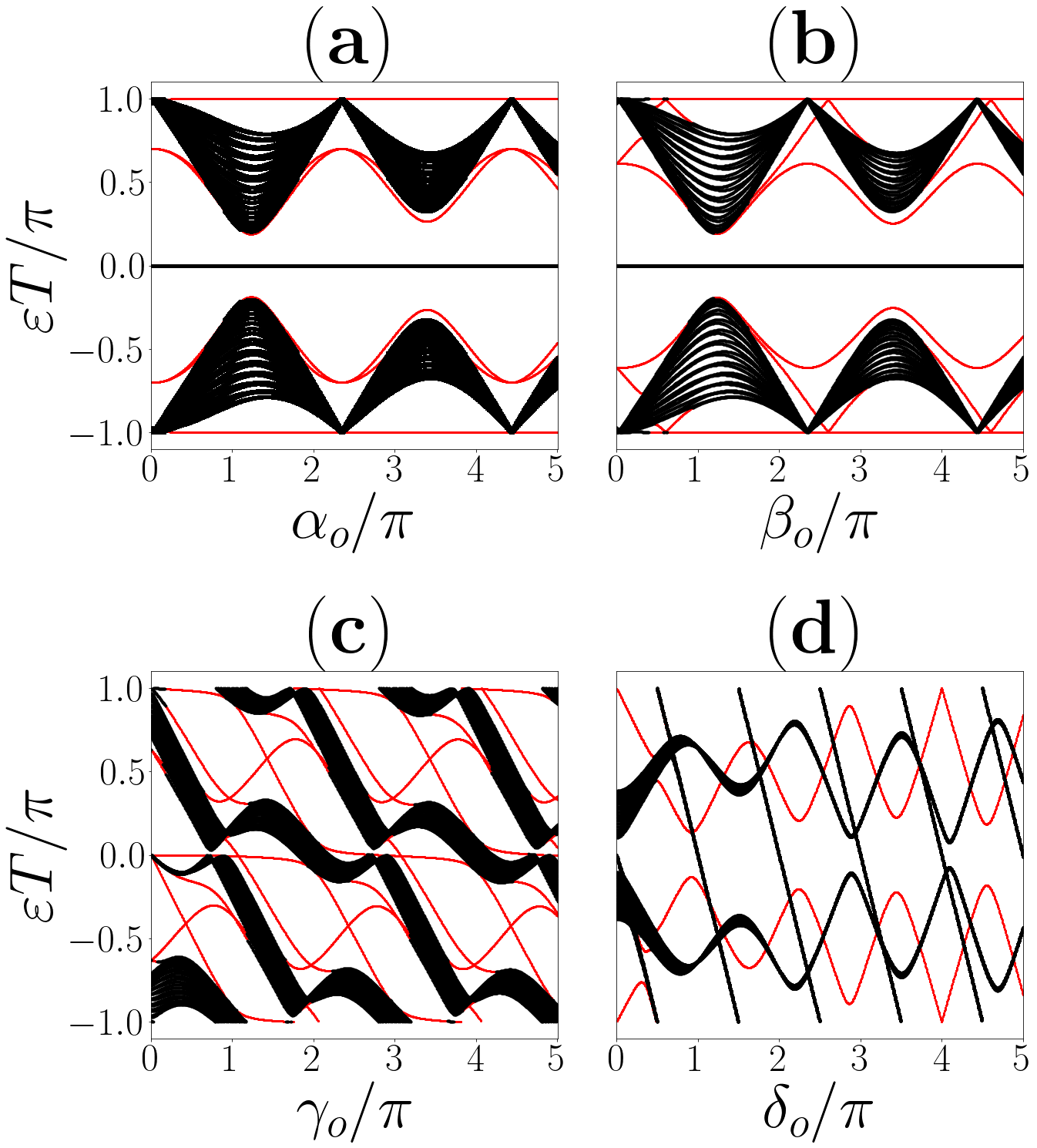}
  \captionof{figure}{Quasienergy spectrum versus the perturbation strengths for Eqs.~\Crefrange{eq:eqa}{eq:eqd} in a system of $N=25$ unit cells at $T=2$. Subplot (a) is at $J_{1}= 1.4 \pi$ and $J_{2}= 0.7\pi$, subplot (b) is at  $J_{1}= 0.7 \pi$ and $J_{2}= 1.4\pi$, subplot (c) is at  $J_{1}= 0.2 \pi$ and $J_{2}= 0.7\pi$, and subplot (d) is at  $J_{1}= 0.1 \pi$ and $J_{2}= 3\pi$. The edge modes are represented in
red.}
  \label{fig:fig13}
\end{figure}    
\end{center}

In Fig.~\ref{fig:fig13}, we plot the system's quasienergy spectrum in the presence of each of the four perturbations introduced in the main text at different sets of parameters and a larger range of perturbation strengths. For all cases, it is clearly observed that the quasienergy structure qualitatively repeats itself as the perturbation strength is increased.

%%%%%%%%%%%%%%%

%\section{Appendix section} 
%\label{app:C}

%%%%%%%%%

\end{document}